\documentclass[notitlepage,nofootinbib,twocolumn,amsmath]{revtex4}

\pdfoutput=1

\usepackage{amsfonts}
\usepackage{amsmath}
\usepackage{amssymb}
\usepackage{aas_macros}
\usepackage{graphicx}
\usepackage{color}
\usepackage{float}
\usepackage{hyperref}
\usepackage[caption = false]{subfig}

\allowdisplaybreaks[1]

\def\lsim{~\rlap{$<$}{\lower 1.0ex\hbox{$\sim$}}}
\def\bsim{~\rlap{$>$}{\lower 1.0ex\hbox{$\sim$}}}

\def\hmpc{\ {\rm {\it h}^{-1}Mpc}}

\def\hmsun{\ {\rm M_\odot/{\it h}}}

\def\hhhmpc{\ {\rm {\it h}^{3}Mpc^{-3}}}

\def\hmmpc{\ {\rm {\it h}Mpc^{-1}}}

\def\la{\langle}
\def\ra{\rangle}

\def\mathbi#1{\textbf{\em #1}}

\def\vk{\mathbi{k}}

\def\vr{\mathbi{r}}

\def\vx{\mathbi{x}}

\def\kmax{k_\text{max}}

\def\apjl{Astrophys.\ J.\ Lett.}

\def\aap{Astron.\ Astrophys.}

\def\apjs{Astrophys.\ J. Supp.}

\definecolor{RedWine}{rgb}{0.743,0,0}
\definecolor{RoyalBlue}{rgb}{0.25,.41,.88}
\definecolor{ForestGreen}{rgb}{.13,.54,.13}
\definecolor{DeepPurple}{rgb}{.72,.18,1}

\begin{document}

\title[Shot noise and biased tracers: a new look at the halo model]{Shot noise and biased tracers: a new look at the halo model}

\author{Dimitry Ginzburg$^1$}

\author{Vincent Desjacques$^{1,2}$}

\email{dvince@physics.technion.ac.il}

\affiliation{$^1$Physics department, Technion, Haifa 3200003, Israel}

\affiliation{$^2$Asher Space Science Institute, Technion, Haifa 3200003, Israel}

\author{Kwan Chuen Chan$^3$}

\affiliation{$^3$Institute of Space Sciences, IEEC-CSIC, Campus UAB, Carrer de Can Magrans, s/n, E-08193 Bellaterra, Barcelona, Spain}

\date{}
\label{firstpage}

\begin{abstract}

Shot noise is an important ingredient to any measurement or theoretical modelling of discrete tracers of the large scale structure.
Recent work has shown that the shot noise in the halo power spectrum becomes increasingly sub-Poissonian at high mass. Interestingly, 
while the halo model predicts a shot noise power spectrum in qualitative agreement with the data, it leads to an unphysical white 
noise in the cross halo-matter and matter power spectrum. In this work, we show that absorbing all the halo model sources of shot 
noise into the halo fluctuation field leads to meaningful predictions for the shot noise contributions to halo clustering statistics
and remove the unphysical white noise from the cross halo-matter statistics. Our prescription straightforwardly maps onto the general
bias expansion, so that the renormalized shot noise terms can be expressed as combinations of the halo model shot noises.
Furthermore, we demonstrate that non-Poissonian contributions are related to volume integrals over correlation functions and their
response to long-wavelength density perturbations. This leads to a new class of consistency relations for discrete tracers, which 
appear to be satisfied by our reformulation of the halo model. 
We test our theoretical predictions against measurements of halo shot noise bispectra extracted from a large suite of numerical 
simulations. Our model reproduces qualitatively the observed sub-Poissonian noise, although it underestimates the magnitude of this 
effect.

\end{abstract}

\maketitle


\section{Introduction}
\label{sec:intro}

Measurements of the distribution of discrete tracers of the large scale structure are affected by cosmic variance and shot noise. 
While sampling variance has been studied thoroughly, shot noise has received less attention because, until recently, the Poissonian 
approximation was sufficient to deal with fairly noisy data. However, ongoing and upcoming surveys like {\small DES}, {\small Euclid}
or {\small LSST} will reduce the statistical uncertainties down to a regime where deviations from Poisson noise become relevant,
and could also be exploited 
(see, e.g., \cite{Seljak:2009af,Hamaus:2010im,Hamaus:2011dq,Hamaus:2012ap,Yoo:2012se,Biagetti:2013sr,Ferraro:2014jba,dePutter:2014lna}).
In fact, deviations from Poisson noise in the clustering of galaxies and clusters have already been reported \cite{Paech:2016hod}.

Early work from \cite{CasasMiranda:2002on,Seljak:2009af} furnished evidence that massive halos have sub-Poissonian noise. This effect 
was thoroughly explored in \cite{Hamaus:2010im}, who studied the eigenvalues of the halo shot noise stochasticity matrix or power 
spectrum $P_{\epsilon\epsilon}(k)$ as a function of halo mass $M$. 
Furthermore, \cite{Hamaus:2010im} proposed a simple analytic expression for the deviation of Poisson noise based on the halo model. 
The halo model \citep[see][for a review]{Cooray:2002dia} provides an analytical framework, which enables us to describe the clustering 
of dark matter and galaxies from the spatial distribution and properties of dark matter halos 
\cite{Seljak:2000gq,Peacock:2000qk,Scoccimarro:2000gm}. However, while \cite{Hamaus:2010im} showed that the halo model reproduces 
reasonably well the properties of the halo shot noise power spectrum $P_{\epsilon\epsilon}(k)$, the model predicts an unphysical white 
noise in the cross halo-matter and matter-matter power spectrum $P_{h\delta}(k)$ and $P_{\delta\delta}(k)$, respectively
\cite{Seljak:2000gq,Cooray:2002dia,Tinker:2004gf,Hamaus:2010im}.

How to cure these inconsistencies is still an open question (see \cite{Scoccimarro:2000gm} for early insights), although
\cite{Smith:2010fh} already noticed that taking into account halo exclusion can cancel the white-noise contribution in the matter power 
spectrum. Ref.\cite{Baldauf:2013hka} argued that the correct, physical explanation for the halo sub-Poissonian noise is halo exclusion 
(which, to some extent, is the reason for nonlinear bias). Using peak theory, which has built-in small-scale exclusion, they showed that 
halo exclusion alone reproduces the measurements of \cite{Hamaus:2010im}. Unfortunately, it is difficult to extract predictions from peak 
theory etc. because small-scale exclusion is, by definition, a highly non-perturbative effect.
An alternative consists in enforcing mass-momentum conservation as in \cite{Schmidt:2015gwz}. This procedure removes, by definition,
all the unphysical terms, yet it does not uniquely constrain the form of the halo shot noise covariances.

Rather than attempting to model halo exclusion from first principles, this paper attempts to retain the simplicity of the halo model. More 
precisely, following a brief overview  in \S\ref{sec:biasPT} of basic theoretical results, we explain in \S\ref{sec:theory} how the 
various sources of shot noise in the halo model should be re-organized such as to preserve the good agreement with the measurements of 
\cite{Hamaus:2010im} and, simultaneously, remove all the unphysical white noises from the theoretical predictions. 
Next, we demonstrate in \S\ref{sec:selfconsistent} that it is possible to map this new halo model onto the general bias expansion, and 
obtain quantitative, unambiguous predictions for the renormalized shot noises. 
Furthermore, we outline in \S\ref{sec:configspace} the consistency relations that exist between shot noise contributions and volume 
integrals over correlation functions.
Finally, in \S\ref{sec:nbody}, we set out to measure bispectrum statistics of the halo shot noise using N-body simulations, and compare
the numerical data with the predictions of our halo model approach. 
After a brief discussion of optimal weights and implementation of halo distribution occupations in \S\ref{sec:discussion}, we conclude in 
\S\ref{sec:conclusions}.

\section{Noise in perturbative bias expansions}
\label{sec:biasPT}

At second order (which is enough for the purpose of this work), the generalized bias expansion takes the form
(see \cite{Desjacques:2016bnm} for a review)
\begin{align}
\label{eq:dhPT}
\delta_h(\vx) &= b_1 \delta(\vx) + b_{\nabla^2\delta}\nabla^2\delta(\vx) + \epsilon_0(\vx)  \\
&\quad + \frac{1}{2} b_2 \delta^2(\vx)  + b_{K^2} \big(K_{ij}^2\big)(\vx) 
+ \epsilon_\delta(\vx)\delta(\vx)
\nonumber \;,
\end{align}
upon neglecting some higher-derivative terms (such as $(\nabla\delta)^2(\vx)$ etc.). 
All the fields on the right-hand side of this equation are non-linear and include fluctuations at all scales, 
except for the long mode, which can be reabsorbed into an effective cosmology. This will be essential to the 
discussion in \S\ref{sec:squeezed}.
The coefficients $b_1$ and $b_2$ are the (Eulerian) linear and quadratic LIMD (local-in-matter-density, see 
\cite{Desjacques:2016bnm}) bias parameters, $b_{\nabla^2\delta}$ is a first-order higher-derivative bias, and 
$b_{K_2}$ is the second-order bias associated with the (traceless) tidal shear tensor $K_{ij}$ (also known as 
tidal shear bias). 
All these bias parameters sensitively depend on the halo mass $M$.
Furthermore, products of fluctuation fields (the equivalent of composite operators) are renormalized, although
our notation does not make it explicit.
The linear term $\epsilon_0(\vx)$ is the usual white noise in the limit $k\to 0$ considered in
\cite{Dekel:1998eq}, whereas the second-order term
$\epsilon_\delta\delta$ represents the response of the halo shot noise to a long-wavelength density perturbation 
\cite{Schmidt:2015gwz}.
No less importantly, the noise fields $\epsilon_0$ and $\epsilon_\delta$ are uncorrelated with density fluctuations,
i.e. $\la\epsilon_0\delta\ra=\la\epsilon_\delta\delta\ra = 0$. However, as we shall see later, they are not fully
specified by their 1-point statistics owing to small-scale exclusion, which introduces correlations on the scale
of dark matter halos. 
At lowest order in perturbation theory, $\epsilon_0$ solely contributes to the halo auto-power spectrum and bispectrum, 
whereas the cross-correlation between $\epsilon_0$ and $\epsilon_\delta$ is relevant to the cross halo-matter 
bispectrum. 

Namely, for the Gaussian initial conditions considered throughout this paper, the halo-matter auto- and cross-bispectra 
which we shall (indirectly) measure in the N-body simulations (cf. Sec.\S\ref{sec:nbody}) are reasonably described 
by the tree-level expressions 
\cite[see in particular][for the shot noise contributions]{1980lssu.book.....P,Matarrese:1997sk,Scoccimarro:2000sn,Pollack:2011xp,Desjacques:2016bnm}
\begin{widetext}
\begin{align}
\label{eq:bispectra}
B_{hhh} &= B_{\epsilon_0\epsilon_0\epsilon_0}^{\{0\}} + b_1^3 B_{\delta\delta\delta}
+ 
\bigg\{b_1^2 \bigg[ b_2 + 2 b_{K^2} \left(\mu_{23}^2-\frac{1}{3}\right)\bigg] P_\text{lin}(k_2)P_\text{lin}(k_3) + \mbox{(2 cyc.)}\bigg\}\\
B_{hh\delta} &= \Big[P_{\epsilon_0\epsilon_\delta}(k_1)+P_{\epsilon_0\epsilon_\delta}(k_2)\Big]P_\text{lin}(k_3)
+ b_1^2 B_{\delta\delta\delta} + 
\bigg\{b_1 \bigg[ b_2 + 2 b_{K^2} \left(\mu_{23}^2-\frac{1}{3}\right)\bigg] P_\text{lin}(k_2)P_\text{lin}(k_3) + \mbox{(2$\longleftrightarrow$3)}\bigg\}
\nonumber \\
B_{h\delta\delta} &= b_1 B_{\delta\delta\delta}+ \bigg[ b_2 + 2 b_{K^2} \left(\mu_{23}^2-\frac{1}{3}\right)\bigg] P_\text{lin}(k_2)P_\text{lin}(k_3) 
\nonumber \;,
\end{align}
\end{widetext}
where $P_\text{lin}(k)$ is the linear density power spectrum, and
\begin{align}
\label{eq:bddd}
B_{\delta\delta\delta} &= 2 \left[\frac{5}{7}+\frac{1}{2}\mu_{12}\bigg(\frac{k_1}{k_2}+\frac{k_2}{k_1}\bigg)+\frac{2}{7}\mu_{12}^2\right] \\
& \qquad \times P_\text{lin}(k_1)P_\text{lin}(k_2) +\mbox{(2 cyc.)} \nonumber
\end{align}
is the tree-level matter bispectrum induced by the gravitational coupling of Fourier modes \cite{1980lssu.book.....P}. 
Here, $\mu_{12}$ is the cosine of the angle between $\vk_1$ and $\vk_2$.

We have not explicitly written the white noise contribution $B_{\epsilon_0\epsilon_0\delta}^{\{0\}}$ to $B_{hh\delta}$, $B_{\epsilon_0\delta\delta}^{\{0\}}$
to $B_{h\delta\delta}$ etc. since they must vanish on the ground that shot noise is uncorrelated with the density field. 
However, as we shall see below, the halo model yields $B_{\epsilon_0\epsilon_0\delta}^{\{0\}}\ne 0$ for instance.
Let us also emphasize that all the noise covariances must converge to their Poissonian expectation in the regime $k_i\gg 1$ 
(i.e. for wavelengths much shorter than the typical size of a halo). For instance, $P_{\epsilon_0\epsilon_0}\to 1/\bar n$, 
$B_{\epsilon_0\epsilon_0\epsilon_0}\to 1/\bar n^2$. One also finds $P_{\epsilon_0\epsilon_\delta}\to 2 b_1/\bar n$ 
\cite{Pollack:2011xp,Schmidt:2015gwz}.
However, in the regime $k_i\ll 1$ of interest for our paper, these noise covariances will generally differ from their Poisson 
expectation. 

We will now illustrate the extent to which the halo model can be applied to infer quantitative estimates in the limit $k_i\to 0$.

\section{Shot-noise in the halo model}
\label{sec:theory}

In the halo model, any polyspectra -- i.e. power spectrum ($N=2$), bispectrum ($N=3$) -- can be written as the sum of 1-halo, 2-halo, ..., 
and $N$-halo contribution. This decomposition can be suitably extended to describe the clustering of dark matter halos themselves 
\cite{Scherrer:1991kk,Smith:2006ne,Hamaus:2010im}. 
For sake of completeness, a selected collection of formulae useful to our analysis is listed in Appendix \S\ref{app:HMformulae}.

After stating our definition of the halo noise we shall measure in simulations, we briefly review in \S\ref{sec:NoiseInPS} how the 
white noise contribution to the halo power spectrum arises from 1-halo terms. 
Next, in \S\ref{sec:NoiseInBS} we scrutinize the shot noise in bispectrum statistics and show that the leading order contributions to 
the noise arise either from 1-halo or 2-halo terms depending on whether one considers auto- or cross- halo-matter bispectra.

\subsection{Halo noise: definition}

We define the halo noise fluctuation field $\epsilon_i$ of a given halo mass bin $i$ as
\begin{align}
\label{eq:defeps}
\epsilon_i(\vk) & \equiv \delta_i(\vk) - b_i \delta(\vk) \\
&= \epsilon_{0i}(\vk) - b_{\nabla^2\delta i}k^2\delta(\vk) \nonumber \\
&\quad +\frac{1}{2} b_{2i} (\delta^2)(\vk)  + b_{K^2i} (K_{ij}^2)(\vk) + (\epsilon_{\delta i}\delta)(\vk) + \dots \nonumber
\end{align}
where $\delta_i(\vk)$ and $\delta(\vk)$ are the Fourier mode of the halo and matter density field, respectively,
$(\delta^2)(\vk)$ is the Fourier transform of $\delta^2$ etc.
This is the quantity whose 2- and 3-point statistics we shall extract from N-body simulations (see \S\ref{sec:nbody}).
Note that, since all the products of fluctuation fields have been renormalized, a constant white noise contribution 
in the limit $k\to 0$ can only stem from a non-vanishing $P_{\epsilon_0\epsilon_0}(k\to 0)$ (which includes the $b_2^2$
1-loop term etc.).

Note that our definition Eq.(\ref{eq:defeps}) of the halo noise is different from that of \cite{Hamaus:2011dq}. While we expect analyzes
with different definitions to agree at large scales, the scale-dependence of the measured halo noise is fairly sensitive to the details 
of the definition adopted.

\subsection{Power spectrum}
\label{sec:NoiseInPS}

The power spectrum of the halo shot noise (or shot noise covariance) is defined as \cite{Hamaus:2010im}
\begin{equation}
P_{\epsilon_i \epsilon_j}(k) \equiv \langle\epsilon_i(\vk)\epsilon_j(-\vk)\rangle' \;,
\end{equation}
where the prime indicates that we have removed a factor of $(2\pi)^3\delta^D(\vk_1+\vk_2)$. The shot noise power spectrum thus reads
\cite{Hamaus:2010im}
\begin{align}
\label{eq:Pee}
P_{\epsilon_i\epsilon_j}(k) &= 
\langle\delta_i\delta_j\rangle(k)-b_i\langle\delta_j\delta\rangle(k) \\ 
&\qquad -b_j\langle\delta_i\delta\rangle(k) +b_i b_j\langle\delta\delta\rangle(k) \nonumber \\
&= P_{ij}(k) - b_i P_{j\delta}(k) - b_j P_{i\delta}(k) + b_i b_j P_{\delta\delta}(k) \nonumber \;.
\end{align}
The contribution from higher-order terms in the bias expansion Eq.(\ref{eq:dhPT}) (such as $b_2\delta^2$) becomes relevant on mildly 
nonlinear scales. Since our focus is on large scales, we shall ignore them in what follows.

In the halo model framework, power spectra decompose into the sum of a 1-halo and a 2-halo term, i.e. 
$P_{XY}(k) \equiv P_{XY}^\text{1H}(k) + P_{XY}^\text{2H}(k)$, where $X$, $Y$ can be any fluctuation field (i.e. matter, 
halo, galaxy etc.). Since 
\begin{align}
P_{ij}^\text{2H}(k) &\propto b_i b_j P_\text{lin}(k) \\
P_{i\delta}^\text{2H}(k) &\propto b_i P_\text{lin}(k) \nonumber \\
P_{\delta\delta}^\text{2H}(k) &\propto P_\text{lin}(k) \nonumber \;,
\end{align}
the 2-halo contributions are negligible relative to the 1-halo terms for $k\ll 1$.
Consequently, shot noise in the limit $k\to 0$ arises exclusively from the 1-halo terms.
For the cross power spectra of halos and matter, these are 
\begin{gather}
P^\text{1H}_{ij}(k) \stackrel{k\to 0}{=} \frac{1}{\bar n_i\bar n_j}\int\!dM\,n(M) \Theta(M,M_i) \Theta(M,M_j)
\equiv \frac{\delta^K_{ij}}{\bar n_i} \nonumber \\
P^\text{1H}_{i\delta}(k) \stackrel{k\to 0}{=} \frac{1}{\bar n_i\bar\rho_m}\int\!dM\,M n(M) \Theta(M,M_i)
\equiv \frac{\overline{M_i}}{\bar\rho_m} \nonumber \\
P^\text{1H}_{\delta\delta}(k) \stackrel{k\to 0}{=} \frac{1}{\bar\rho_m^2}\int\!dM\,M^2 n(M)
\equiv \frac{\la nM^2\ra}{\bar\rho_m^2}
\label{eq:P1H}
\end{gather}
upon taking the large-scale limit for the Fourier transform $u$ of the halo profile, i.e. $u(k\to 0|M)=1$. 
Here, $P_{i\delta}$ denotes the cross-power spectrum between the dark matter and the $i$th halo density field $\delta_i$,
while $P_{ij}$ is the cross-power spectrum between the $i$th and $j$th halo fluctuation field $\delta_i$ and $\delta_j$. 
The latter formally reads
\begin{equation}
\label{eq:delta_i}
\delta_i(\vx) = \frac{n_i(\vx)}{\bar n_i}-1 \;.
\end{equation} 
Furthermore, $\Theta(M,M_i)$ is a combination of step functions which returns unity when $M$ lies within the mass bin 
centered at $M_i$, and zero otherwise. 
$n(M)$ is the number density of halos in the mass range $[M,M+dM]$ or halo mass function 
(and thus has units of inverse Mass$\cdot$Volume), so that
\begin{equation}
\bar n_i = \int\!dM\,n(M)\Theta(M,M_i)
\end{equation}
is the number density of halos in the $i$th bin. Note that our notation distinguishes between averages like
\begin{equation}
\overline{M_i^kb_i^l}\equiv \frac{1}{\bar n_i}\int\!dM\,M^k n(M) b_1^l(M) \Theta(M,M_i)
\end{equation}
over a single mass bin, which are indicated with an overline, and averages like
\begin{equation}
\la n M^k b_1^l\ra\equiv \int\!dM\,M^k n(M) b_1^l(M)
\end{equation}
over all halos, which are denoted with the brackets $\la ... \ra$. 
In practice, we choose a finite lower bound (e.g. equal to the mass of one dark matter particle in the numerical simulations) 
to avoid numerical instabilities at low mass. We have found that numerical estimates of 
$\la n M^k b_1^l\ra$ are not very sensitive to the lower limit of the integral.
Finally, $\delta^K_{ij}$ is a Kronecker symbol, i.e. $\delta_{ij}^K=1$ if $i=j$ and zero otherwise.

Substituting the 1-halo contributions Eq.(\ref{eq:P1H}) into Eq.(\ref{eq:Pee}), the $k\to 0$ limit of halo noise power
spectrum predicted by the halo model reads \cite{Hamaus:2010im}:
\begin{align}
\label{eq:Pe0e0}
P_{\epsilon_i\epsilon_j}(k\to 0) & \equiv P_{\epsilon_{0i}\epsilon_{0j}}^{\{0\}} \\
&= \frac{1}{\bar n_i}\delta_{ij}^K - b_i\frac{\overline{M_j}}{\bar\rho_m}-b_j\frac{\overline{M_i}}{\bar\rho_m}
+b_i b_j \frac{\la n M^2\ra}{\bar\rho_m^2} \nonumber \;.
\end{align}
For the same halo mass bin ($i=j$), we get
\begin{equation}
\label{eq:Pe0e0ii}
P_{\epsilon_0\epsilon_0}^{\{0\}} =\frac{1}{\bar n}\bigg\{
1 - 2 b_1\left(\frac{\bar n\overline{M}}{\bar\rho_m}\right)+b_1^2 \left(\frac{\bar n\la n M^2\ra}{\bar\rho_m^2}\right)
\bigg\} 
\end{equation}
Interestingly, Eq.(\ref{eq:Pe0e0}) turns out to be in very good agreement with the numerical data as shown in \cite{Hamaus:2010im}, 
although some of the halo model predictions are unphysical. 
Namely, there is no shot noise in the cross halo-mass power spectrum as this is nothing but the average density profile. Therefore, 
we should have $P_{i\delta}^\text{1H}\to 0$ in the limit $k\to 0$. Nonetheless, the halo model assumes
\begin{equation}
\la \epsilon_i \delta\ra  \to \frac{\overline{M_i}}{\bar\rho_m} \ne 0 \;,
\end{equation}
where $\bar\rho_m$ is the average matter density in the Universe.

Similarly, we would also expect $P_{\delta\delta}^\text{1H}\to 0$ in the ``thermodynamic'' limit $m_\text{DM}\to 0$, 
where $m_\text{DM}$ is the mass of the dark matter particles. 
However, the halo model predicts $P_{\delta\delta}^\text{1H}\to \la nM^2\ra/\bar\rho_m^2$, a value noticeably larger 
than the Poisson $1/\bar n_\text{DM}$ (where $\bar n_\text{DM}$ is the number density of DM particles) observed in 
numerical simulations \cite{Cooray:2002dia,Smith:2002dz,Crocce:2007dt,Hamaus:2010im}.
These are only two among infinitely many examples for which the halo model does not yield physically consistent 
results.

Finally, standard implementations of the halo model ignore the possibility that the shot noise may not be Poissonian 
in the halo power spectrum, although there is evidence that this is not the case 
\cite{CasasMiranda:2002on,Seljak:2009af,Hamaus:2010im}.
We will show in \S\ref{sec:selfconsistent} how these issues can be resolved all at once.

\subsection{Bispectrum}
\label{sec:NoiseInBS}

Let us now turn to the bispectrum.
Since we shall need cross-bispectra of different fluctuation fields, it is convenient to define the bispectrum of 
three fluctuation fields $X(\vk_1)$, $Y(\vk_2)$ and $Z(\vk_3)$ as 
\begin{equation}
\label{eq:Bdef}
B_{XYZ}(k_1,k_2,k_3) \equiv \langle X(\vk_1) X(\vk_2) Y(\vk_3)\rangle ' \;,
\end{equation}
where, again, the `` ' ``attached to Fourier space correlators signifies that we have removed a factor of 
$(2\pi)^3 \delta^D\!(\vk_1+\vk_2+\vk_3)$ owing to the invariance under translations ($\delta^D$ is the Dirac distribution).
In the halo model, bispectra can be written as a sum of a 1-, 2- and 3-halo term.
To avoid clutter, we will often omit the explicit dependence of $B$ on the wavenumbers $k_i$. However, one should bear in
mind that the wavenumbers are always ordered with the fluctuation fields as in Eq.(\ref{eq:Bdef}).

Unlike power spectra, for which tree-level shot noise contributions can arise only through 1-halo terms, for bispectra 
these can arise either through the 1-halo or 2-halo terms. 

To illustrate this point, let us begin with the bispectrum 
$B_{\epsilon_i\epsilon_j\epsilon_k}=\la\epsilon_i(\vk_1)\epsilon_j(\vk_2)\epsilon_k(\vk_3)\ra'$ 
of the halo noise fluctuation field $\epsilon_i$,
\begin{widetext}
\begin{align}
B_{\epsilon_i\epsilon_j\epsilon_k} &= \langle\delta_i\delta_j\delta_k\rangle'
-\Big[b_i\langle\delta_j\delta_k\delta\rangle' + \mbox{(2 cyc.)}\Big] 
+\Big[b_i b_j\langle\delta\delta\delta_k\rangle' + \mbox{(2 cyc.)}\Big]
- b_i b_j b_k\langle\delta\delta\delta\rangle' \nonumber \\
&= \Big. B_{ijk} - b_i B_{\delta jk} - b_j B_{i\delta k}  - b_k B_{ij\delta}
+  b_k b_j B_{i\delta\delta} + b_i b_k B_{\delta j\delta}  + b_i b_j B_{\delta\delta k} 
- b_i b_j b_k B_{\delta\delta\delta} \;,
\label{eq:Beee}
\end{align}
\end{widetext}
where, for shorthand convenience, $B_{ijk}$, $B_{ij\delta}$ etc. denote the halo bispectrum, the cross halo-matter bispectra 
etc. evaluated for the triplet of wavenumbers $(k_1,k_2,k_3)$. The subscripts are ordered such that the first carries momentum 
$\vk_1$ and the third $\vk_3$. 

$B_{\epsilon_i\epsilon_j\epsilon_k}$ contributes a white noise in the limit $k_i\to 0$, which we shall denote by 
$B_{\epsilon_i\epsilon_j\epsilon_k}^{\{0\}}$ (this corresponds to $B_\epsilon^{\{0\}}$ in the notation of \cite{Desjacques:2016bnm}). 
From the point of view of the halo model, since the various 2- and 3-halo contributions $B^\text{2H}$ and $B^\text{3H}$ are 
proportional to the linear power spectrum $P_\text{lin}$ and, thereby, vanish in the limit $k_i\to 0$. 
The only terms relevant to  $B_{\epsilon_i\epsilon_j\epsilon_k}^{\{0\}}$ thus are the 1-halo bispectra. Explicit expressions at finite
wavenumber can be found in Appendix \S\ref{app:HMformulae}. In the low-$k$ limit, these become
\begin{widetext}
\begin{gather}
B^\text{1H}_{ijk}(k_1,k_2,k_3) \stackrel{k_i\to 0}{=} \frac{1}{\bar n_i\bar n_j\bar n_k}
\int\!dM\,n(M) \Theta(M,M_i) \Theta(M,M_j) \Theta(M,M_k)
\equiv \frac{1}{\bar n_i^2}\delta^K_{ijk}\nonumber \\
B^\text{1H}_{ij\delta}(k_1,k_2,k_3) \stackrel{k_i\to 0}{=} \frac{1}{\bar n_i\bar n_j\bar\rho_m}
\int\!dM\,M n(M) \Theta(M,M_i) \Theta(M,M_j) 
\equiv \frac{\overline{M_i}}{\bar n_i\bar\rho_m}\delta^K_{ij}\nonumber \\
B^\text{1H}_{i\delta\delta}(k_1,k_2,k_3) \stackrel{k_i\to 0}{=} \frac{1}{\bar n_i\bar\rho_m^2}
\int\!dM\,M^2 n(M) \Theta(M,M_i) 
\equiv \frac{\overline{M_i^2}}{\bar\rho_m^2} \nonumber \\
B^\text{1H}_{\delta\delta\delta}(k_1,k_2,k_3) \stackrel{k_i\to 0}{=}
\frac{1}{\bar\rho_m^3}\int\!dM\,M^3 n(M) \equiv \frac{\la nM^3\ra}{\bar\rho_m^3}
\label{eq:B1H}
\end{gather}
\end{widetext}
Note that, for a halo abundance $\bar n_i\ll 1$, each of these 1-halo term is suppressed relative to $B^\text{1H}_{ijk}$ by a 
(dimensionless) factor of $(\bar n \overline{M}/\bar\rho_m)^k$, where $k$ is the number of density field $\delta$. 
Substituting the halo model predictions Eq.(\ref{eq:B1H}) into the definition Eq.(\ref{eq:Beee}) of the halo noise bispectrum, 
we obtain
\begin{align}
\label{eq:Be0e0e0}
B_{\epsilon_i\epsilon_j\epsilon_k}(k_i\to 0) & \equiv B_{\epsilon_{0i}\epsilon_{0j}\epsilon_{0k}}^{\{0\}}\\
&= - b_i b_j b_k \frac{\la n M^3\ra}{\bar\rho_m^3}+\bigg[b_i b_j \frac{\overline{M_k^2}}{\bar\rho_m^2}+ \mbox{(2 cyc.)}\bigg] 
\nonumber \\
& \qquad - \bigg[b_i\left(\frac{\overline{M_j}}{\bar n_j\bar\rho_m}\right)\delta_{jk}^K + \mbox{(2 cyc.)}\bigg] 
+ \frac{\delta_{ijk}^K}{\bar n_i^2} \nonumber \;.
\end{align}
The magnitude of $B_{\epsilon_{0i}\epsilon_{0j}\epsilon_{0k}}^{\{0\}}$ strongly depends on the number of identical noise field. 
When two or three noise fluctuation fields are different, it is suppressed by one or two factors of $\bar n \overline{M}/\bar\rho_m$
relative to the Poisson expectation $1/\bar n^2$. By contrast, when the three noise fluctuation fields are identical, we find
\begin{align}
\label{eq:Be0e0e0iii}
B_{\epsilon_0\epsilon_0\epsilon_0}^{\{0\}} &=\frac{1}{\bar n^2}
\bigg\{1 - 3 b_1\left(\frac{\bar n \overline{M}}{\bar\rho_m}\right) \\ 
&\qquad +3b_1^2 \left(\frac{\bar n^2 \overline{M^2}}{\bar\rho_m^2}\right) 
- b_1^3 \left(\frac{\bar n^2\la n M^3\ra}{\bar \rho_m^3}\right) \bigg\} 
\nonumber \;,
\end{align}
where all the parentheses within the curly brackets are dimensionless quantities

Eq.(\ref{eq:Be0e0e0iii}) furnishes a prediction for the zero-point amplitude of the shot noise bispectrum 
$B^{\{0\}}_{\epsilon_0\epsilon_0\epsilon_0}$.
As we shall see in \S\ref{sec:nbody}, this prediction broadly agrees with measurements of the halo shot noise bispectrum 
extracted from a large suite of numerical simulations. 
However, Eq.(\ref{eq:B1H}) shows that the halo model predicts, among others, a constant white noise contribution 
$B_{ii\delta}\to (\overline{M}_i/\bar n_i\bar\rho_m)\ll\bar n_i^{-2}$ in the limit $k_i\to 0$. 
Like for the cross halo-matter power spectrum $P_{i\delta}$, this contribution is unphysical because the matter fluctuation 
field is devoid of shot noise in the thermodynamic limit $m_\text{DM}\to 0$. Therefore, shot noise in $B_{ij\delta}$ can 
only arise from second- or higher-order contributions to $\epsilon_i$ which, at lowest order, are proportional to
$P_\text{lin}(k_i)$. 
In the halo model, terms linear in $P_\text{lin}$ arise from 2-halo terms. 

The relevant 2-halo contributions are listed in Appendix \S\ref{app:HMformulae}. In the low-$k$ limit, they are given by 
\begin{widetext}
\begin{align}
B^\text{2H}_{ij\delta}(k_1,k_2,k_3) &\stackrel{k_i\to 0}{=} 
\frac{1}{\bar n_i\bar n_j\bar\rho_m}\int\!dM\,\Theta(M,M_i) n(M) b_1(M)\int\!dM'\,\Theta(M',M_j)M'n(M')  b_1(M') P_\text{lin}(k_1)
\nonumber \\
&\qquad + \frac{1}{\bar n_i\bar n_j\bar\rho_m}\int\!dM\,\Theta(M,M_j) n(M) b_1(M)\int\!dM'\,\Theta(M',M_i)M' n(M') b_1(M') P_\text{lin}(k_2)
\nonumber \\
& \qquad + \frac{1}{\bar n_i\bar n_j\bar\rho_m}\int\!dM\,M n(M) b_1(M)\int\!dM'\,\Theta(M',M_i)\Theta(M',M_j) n(M') b_1(M') P_\text{lin}(k_3)
\nonumber \\
&\quad\equiv b_i \frac{\overline{M_jb_j}}{\bar\rho_m}P_\text{lin}(k_1)+b_j\frac{\overline{M_ib_i}}{\bar\rho_m}P_\text{lin}(k_2) 
+\frac{b_i}{\bar n_i}\delta_{ij}^K P_\text{lin}(k_3) \nonumber \\
B^\text{2H}_{i\delta\delta}(k_1,k_2,k_3) &\stackrel{k_i\to 0}{=} 
\frac{1}{\bar n_i\bar\rho_m^2}\int\!dM\,\Theta(M,M_i) n(M) b_1(M)\int\!dM'\,(M')^2 n(M')b_1(M') P_\text{lin}(k_1)
\nonumber \\
&\qquad + \frac{1}{\bar n_i\bar\rho_m^2}\int\!dM\,M n(M) b_1(M)\int\!dM'\,\Theta(M',M_i)M'n(M')b_1(M')
\Big[P_\text{lin}(k_2)+P_\text{lin}(k_3)\Big] 
\nonumber \\
&\quad\equiv b_i \frac{\la n M^2 b_1\ra}{\bar\rho_m^2} P_\text{lin}(k_1)+\frac{\overline{M_ib_i}}{\bar\rho_m}
\Big[P_\text{lin}(k_2) +P_\text{lin}(k_3)\Big] \nonumber \\
B^\text{2H}_{\delta\delta\delta}(k_1,k_2,k_3) &\stackrel{k_i\to 0}{=} 
\frac{1}{\bar\rho_m^3}\int\!dM\,M n(M) b_1(M)\int\!dM'\,\big(M'\big)^2 n(M')b_1(M')P_\text{lin}(k_3) + \mbox{(2 cyc.)}
\nonumber \\
&\quad\equiv \frac{\la n M^2 b_1\ra}{\bar\rho_m^2}P_\text{lin}(k_3) + \mbox{(2 cyc.)}\;,
\label{eq:B2H}
\end{align}
\end{widetext}
where the wavemode $k_3$ is always assigned to the last (density) fluctuation field. 
Furthermore, $\overline{M_ib_i}$ designates the average of the product $M\cdot b_1$ across one halo bin.
To obtain the last equalities, we have extensively applied the relation
\begin{equation}
\int\!dM\,M n(M) b_1(M) \equiv \bar\rho_m \;,
\end{equation}
which follows from mass conservation and the peak-background split relation for the linear halo bias 
(see \cite{Desjacques:2016bnm} and references therein),
\begin{equation}
\label{eq:PBSb1}
b_1 = \frac{\bar\rho_m}{\bar n}\frac{\partial\bar n}{\partial\bar\rho_m} \;.
\end{equation}
Substituting the 1-halo and 2-halo expressions (\ref{eq:B1H}) and (\ref{eq:B2H}) into the cross-bispectrum 
$B_{\epsilon_i\epsilon_j\delta}(k_1,k_2,k_3)\equiv \langle\epsilon_i(\vk_1)\epsilon_j(\vk_2)\delta(\vk_3)\rangle '$
of halo noise and matter fluctuation fields,
\begin{equation}
\label{eq:Beed}
B_{\epsilon_i\epsilon_j\delta} = \Big. B_{ij\delta}-b_i B_{\delta j \delta}-b_j B_{i\delta\delta}+b_i b_j B_{\delta\delta\delta} \;,
\end{equation}
a naive application of the halo model yields
\begin{align}
\label{eq:Be0e0ed}
B_{\epsilon_i\epsilon_j\delta}(k_i\to 0) &\equiv B_{\epsilon_{0i}\epsilon_{0j}\delta}^{\{0\}}  \\
&\qquad + \Big(P_{\epsilon_{0i}\epsilon_{\delta j}}^{\{0\}}+P_{\epsilon_{0j}\epsilon_{\delta i}}^{\{0\}}\Big) P_\text{lin}(k_3) 
\nonumber \;,
\end{align}
where 
\begin{equation}
\label{eq:unphysicalBeed}
B_{\epsilon_{0i}\epsilon_{0j}\delta}^{\{0\}} = \frac{\overline{M_i}}{\bar n_i\bar\rho_m}\delta_{ij}^K-b_i\frac{\overline{M_j^2}}{\bar\rho_m^2}
-b_j\frac{\overline{M_i^2}}{\bar\rho_m^2}+b_i b_j\frac{\la nM^3\ra}{\bar\rho_m^3}
\end{equation}
is the white noise piece arising from the 1-halo terms. 
For the 2-halo piece, some cancellations occur and we are left with
\begin{align}
\label{eq:Pe0ed}
P_{\epsilon_{0i}\epsilon_{\delta j}}^{\{0\}}+P_{\epsilon_{0j}\epsilon_{\delta i}}^{\{0\}} &= 
\frac{b_i}{\bar n_i}\delta_{ij}^K-b_i \frac{\overline{M_jb_j}}{\bar\rho_m} - b_j \frac{\overline{M_i b_i}}{\bar\rho_m} \\
&\qquad +b_i b_j\frac{\la n M^2 b_1\ra}{\bar\rho_m^2} \nonumber \;.
\end{align}
This is the amplitude proportional to $P_\text{lin}(k_3)$ in the limit $k_i\ll 1$.

In the next Section, we shall argue that $B_{\epsilon_{0i}\epsilon_{0j}\delta}^{\{0\}}\equiv 0$, in contrast to Eq.(\ref{eq:unphysicalBeed}), 
is the correct answer. However, we will also see that the prediction for the amplitude 
$P_{\epsilon_{0i}\epsilon_{\delta j}}^{\{0\}}+P_{\epsilon_{0j}\epsilon_{\delta i}}^{\{0\}}$ of the term proportional to $P_\text{lin}(k_3)$ obtained
from a brute force application of the halo model is physically sound. 
This amplitude is generally negative for $i\ne j$ whereas, in the particular case $i=j$, we find
\begin{equation}
\label{eq:Pe0edii}
P_{\epsilon_0\epsilon_\delta}^{\{0\}} = \frac{b_1}{2\bar n}\bigg\{1 -2 \left(\frac{\bar n\overline{M b_1}}{\bar\rho_m}\right)
+b_1\left(\frac{\bar n \la n M^2 b_1\ra}{\bar\rho_m^2}\right)\bigg\}
\end{equation}
The first term in the parenthesis yields the Poisson expectation (already derived in \cite{Pollack:2011xp,Schmidt:2015gwz}), which is 
expected to be valid in the limit $k\gg 1$ only. The other terms are the non-Poissonian corrections that arise from halo exclusion. 
Note that the non-Poissonian correction in Eq.(\ref{eq:Pe0edii}) differ from that in Eq.(\ref{eq:Pe0e0ii}) due to the position of a 
factor of $b_1$ in the third term.

The generalization of these calculations to higher-order polyspectra is straightforward. As a general rule, a naive application of the 
halo model will predict physically consistent noise covariances so long as they do not involve any density fluctuation field, and fail
in all other cases.

\section{Self-consistent treatment of  shot noise in the halo model}
\label{sec:selfconsistent}

In the previous Section, we have shown that the halo model can used to compute shot noise contribution beyond the constant white noise
term scrutinized in the study of \cite{Hamaus:2010im}. However, as we have already emphasized, some of the halo model predictions are 
unphysical. For bispectra for instance, there should not be any constant white noise in the limit $k_i\to 0$ unless the bispectrum is 
computed for three halo fluctuation fields. How can we avoid these inconsistencies and simultaneously retain the halo model predictions 
for $P_{\epsilon_0\epsilon_0}^{\{0\}}$ or $B_{\epsilon_0\epsilon_0\epsilon_0}^{\{0\}}$ which, as we shall see in \S\ref{sec:nbody}, are in 
reasonably good agreement with the data ?

To address this question, we begin by writing down a perturbative expansion for both the halo and matter overdensity fields that 
reproduces the ``original'' halo model predictions Eqs.~(\ref{eq:P1H}), (\ref{eq:B2H}) and (\ref{eq:B1H}). In light of the bias expansion
Eq.(\ref{eq:dhPT}), we try to following ansatz:
\begin{align}
\Big.\delta_i(\vx) &= \big(b_i + \tilde\epsilon_{\delta i}(\vx)\big) \delta(\vx) + \tilde \epsilon_{0i}(\vx) \nonumber \\
\Big.\delta_m(\vx) &= \big(1 + \tilde\epsilon_{\delta m}(\vx)\big) \delta(\vx) + \tilde \epsilon_{0m}(\vx)\;,
\label{eq:oldHMnoise}
\end{align}
where we have ignored second- and higher-order terms.
Here, $\delta$ should be interpreted as the noise-free, nonlinear density field, whereas $\tilde\epsilon_{\delta m}$ and $\tilde \epsilon_{0m}$ 
are the matter equivalents to the lowest order halo shot noise terms. In the halo model, they do not vanish even in the limit $m_\text{DM}\to 0$
considered throughout our calculations.
For clarity, we use tilde in order to distinguish the halo shot noise contributions from the renormalized shot noise terms which appear 
in Eq.(\ref{eq:dhPT}). 

To assess the extent to which Eq.(\ref{eq:oldHMnoise}) reproduces the halo model predictions in the low-$k$ limit, we compute the large-scale 
2-point and 3-point covariances of the noise fields $\tilde\epsilon_{0i}$, $\tilde\epsilon_{\delta i}$, $\tilde\epsilon_{0m}$ and 
$\tilde\epsilon_{0\delta}$ using our ansatz for the halo and matter fluctuation field $\delta_h(\vx)$ and $\delta_m(\vx)$ (which replaces 
$\delta(\vx)$ for this consistency check), respectively. 
We consider first the 2-point covariances.
The calculation of $\la\delta_h\delta_h\ra$, $\la\delta_h\delta_m\ra$ and $\la\delta_m\delta_m\ra$ in the limit $k\to 0$ is straightfoward. 
Upon identifying our results with Eq.(\ref{eq:P1H}), the following noise power spectra must satisfy:
\begin{gather}
P_{\tilde\epsilon_{0i}\tilde\epsilon_{0j}}^{\{0\}} = \frac{\delta_{ij}^K}{\bar n_i} \;,\quad 
P_{\tilde\epsilon_{0i}\tilde\epsilon_{0m}}^{\{0\}} = \frac{\overline{M_i}}{\bar\rho_m}\;,
\label{eq:2ptnoise1} \\
P_{\tilde\epsilon_{0m}\tilde\epsilon_{0m}}^{\{0\}} = \frac{\la\bar n M^2\ra}{\bar\rho_m^2}\nonumber \;.
\end{gather}
Similarly, the 2-halo contribution to the various cross halo-matter bispectra $\la\delta_h\delta_h\delta_m\ra$ etc. in the limit $k_i\ll 1$, 
Eq.(\ref{eq:B2H}), constrain another set of 2-point covariances. 
Namely, terms proportional to $P_\text{lin}(k_3)$ arise from 4-point correlators involving two shot noise and two density fields (5-point 
correlators of the form $\la(\epsilon_X\delta)(\vk_1)(\epsilon_Y\delta)(\vk_2)\epsilon_Z(\vk_3)\ra$ return a loop).
In the limit $k_i\to 0$, the relevant shot noise contributions are:
\begin{widetext}
\begin{align}
B_{mmm}(k_1,k_2,k_3) & 
\supset \Big[ P_{\tilde\epsilon_{\delta m}\tilde\epsilon_{0m}}(k_1) + P_{\tilde\epsilon_{\delta m}\tilde\epsilon_{0m}}(k_2)\Big] P_\text{lin}(k_3) 
+ \mbox{(2 cyc.)} \\
B_{imm}(k_1,k_2,k_3) &
\supset b_i \Big[P_{\tilde\epsilon_{\delta m}\tilde\epsilon_{0m}}(k_2) + P_{\tilde\epsilon_{\delta m}\tilde\epsilon_{0m}}(k_3)\Big] P_\text{lin}(k_1)
+ \Big[P_{\tilde\epsilon_{\delta i}\tilde\epsilon_{0 m}}(k_3) + P_{\tilde\epsilon_{0 i}\tilde\epsilon_{\delta m}}(k_1)\Big]P_\text{lin}(k_2) 
\nonumber \\
& \qquad + \Big[P_{\tilde\epsilon_{0 i}\tilde\epsilon_{\delta m}}(k_1) + P_{\tilde\epsilon_{\delta i}\tilde\epsilon_{0 m}}(k_2)\Big]P_\text{lin}(k_3)
\nonumber \\
B_{ijm}(k_1,k_2,k_3) &
\supset b_i \Big[P_{\tilde\epsilon_{0 j}\tilde\epsilon_{\delta m}}(k_2) + P_{\tilde\epsilon_{\delta j}\tilde\epsilon_{0 m}}(k_3)\Big]P_\text{lin}(k_1)
+ b_j \Big[P_{\tilde\epsilon_{\delta i}\tilde\epsilon_{0 m}}(k_3) + P_{\tilde\epsilon_{0 i }\tilde\epsilon_{\delta m}}(k_1)\Big]P_\text{lin}(k_2)
\nonumber \\
&\qquad + \Big[P_{\tilde\epsilon_{0 i}\tilde\epsilon_{\delta j}}(k_1) + P_{\tilde\epsilon_{\delta i}\tilde\epsilon_{0 j}}(k_2)\Big]P_\text{lin}(k_3)
\nonumber \;.
\end{align}
\end{widetext}
Identifying the above expressions with Eq.(\ref{eq:B2H}), the following combination of zero-lag noise power spectra must satisfy
\begin{gather}
P_{\tilde\epsilon_{\delta m}\tilde\epsilon_{0m}}^{\{0\}} = \frac{\la n M^2 b_1\ra}{2\bar\rho_m^2} \;, \quad
P_{\tilde\epsilon_{\delta i}\tilde\epsilon_{0m}}^{\{0\}} + P_{\tilde\epsilon_{0 i}\tilde\epsilon_{\delta m}}^{\{0\}} = 
b_i\frac{\overline{M_i}}{\bar\rho_m} \;, \nonumber \\
P_{\tilde\epsilon_{\delta i}\tilde\epsilon_{0 j}}^{\{0\}} + P_{\tilde\epsilon_{0 i}\tilde\epsilon_{\delta j}}^{\{0\}} =
\frac{b_i}{\bar n_i} \delta_{ij}^K \label{eq:2ptnoise2} \;.
\end{gather} 
Finally, some of the 3-point noise covariances can be read off from Eqs.(\ref{eq:B1H}). We find:
\begin{gather}
B_{\tilde\epsilon_{0i}\tilde\epsilon_{0j}\tilde\epsilon_{0k}}^{\{0\}} = \frac{1}{\bar n_i^2}\delta_{ijk}^K \;, \quad
B_{\tilde\epsilon_{0i}\tilde\epsilon_{0j}\tilde\epsilon_{0m}}^{\{0\}} = \frac{\overline{M_i}}{\bar n_i\bar\rho_m}\delta_{ij}^K
\label{eq:3ptnoise} \\
B_{\tilde\epsilon_{0i}\tilde\epsilon_{0m}\tilde\epsilon_{0m}}^{\{0\}} = \frac{\overline{M_i^2}}{\bar\rho_m^2} \;,\quad
B_{\tilde\epsilon_{0m}\tilde\epsilon_{0m}\tilde\epsilon_{0m}}^{\{0\}} = \frac{\la n M^3\ra}{\bar\rho_m^3} 
\nonumber \;.
\end{gather}
In particular, $B_{ijm}$ exhibits a constant white noise in the low-$k$ because $B_{\tilde\epsilon_{0i}\tilde\epsilon_{0j}\tilde\epsilon_{0m}}^{\{0\}}\ne 0$.
All this suggests that the halo model indeed assumes that the halo and matter overdensity be described by perturbative expansions of the 
form Eq.(\ref{eq:oldHMnoise}). 

In order to remedy the unphysical, large-scale behaviour of all the 1-halo terms involving at least one density field, and simultaneously retain 
the halo model predictions for $P_{\epsilon_{0i}\epsilon_{0j}}^{\{0\}}$, $P_{\epsilon_{0i}\epsilon_{\delta j}}^{\{0\}}$ etc., we argue that the halo model 
perturbative expansion Eq.(\ref{eq:oldHMnoise}) should be reorganized such that
\begin{align}
  \label{eq:newHMnoise}
\Big.\delta_i(\vx) &= \big(b_i + \tilde\epsilon_{\delta i}(\vx) - b_i \tilde\epsilon_{\delta m}(\vx)\big) \delta(\vx)  \\
&\qquad \Big. + \tilde\epsilon_{0i}(\vx) - b_i\tilde\epsilon_{0m}(\vx)
\nonumber \\
\Big.\delta_m(\vx) &= \delta(\vx) \nonumber \;.
\end{align}
Here, $\delta(\vx)$ is the nonlinear density field as in the pertubative bias expansion Eq.(\ref{eq:dhPT}).
At the tree-level considered in this paper, this ansatz clearly ensures that any correlator involving at least one matter fluctuation field does 
not exhibit a constant white noise in the limit $k_i\to 0$ (i.e., the right-hand side of Eq.(\ref{eq:unphysicalBeed}) now vanishes). 
In particular,
\begin{align}
\bigg. P_{i\delta}(k) &\stackrel{k\ll 1}{=} b_i P_\text{lin}(k) \nonumber \\
\bigg. P_{\delta\delta}(k) &\stackrel{k\ll 1}{=} P_\text{lin}(k)
\end{align}
in the thermodynamics limit  $m_\text{DM}\to 0$. 
At the same time, Eqs.(\ref{eq:Pe0e0}), (\ref{eq:Be0e0e0}) and (\ref{eq:Pe0ed}) are reproduced. This way, the halo model can be used to
consistently predict the shot noise contributions that arise from the perturbative bias expansion Eq.(\ref{eq:dhPT}) since we can identify
the renormalized shot noises $\epsilon_{0}$ and $\epsilon_{\delta}$ with the halo model shot noises according to
\begin{align}
\label{eq:newPTnoise}
\Big.\epsilon_{0i}(\vx) &\equiv \tilde\epsilon_{0i}(\vx) - b_i \tilde \epsilon_{0m}(\vx) \\
\Big.\epsilon_{\delta i}(\vx) & \equiv \tilde\epsilon_{\delta i}(\vx) - b_i \tilde\epsilon_{\delta m}(\vx) \nonumber \;.
\end{align}
Eqs.(\ref{eq:newHMnoise}) and (\ref{eq:newPTnoise}) are the main results of this Section. They provide the building blocks for a self-consistent
calculation of halo clustering statistics based on the halo model. We shall explore the extent to which our prescription generalizes to higher
order in future work. For the moment, let us emphasize two important points.

Firstly, while the Poisson noise piece $\tilde\epsilon_{0i}(\vx)$ has non-vanishing correlations at one-point only, e.g. 
$\la\tilde\epsilon_{0i}(\vx_1)\tilde\epsilon_{0j}(\vx_2)\ra\sim \delta^D\!(\vx_1-\vx_2)$, the remaining auto- and cross-correlations are generally
non-zero at separations $|\vx_1-\vx_2|\lesssim R_\text{vir}$, where $R_\text{vir}$ is the halo virial radius. This should be interpreted as a
manifestation of halo exclusion, which introduces correlations at small scales.
Therefore, the physical interpretation of Eq.(\ref{eq:newHMnoise}) is that, while the matter distribution has zero shot noise in the limit
$m_\text{DM}\to 0$ and, thus, satisfies the mass-momentum conservation, the halo shot noise is a combination of Poisson sampling noise and exclusion
effects.

Secondly, a natural extension of our predictions to finite wavenumber would consist in replacing the covariances of the halo shot noises 
$\tilde\epsilon_{0i}$, $\tilde\epsilon_{\delta i}$ etc. by the halo model predictions at finite $k$. The 1-halo and 2-halo power spectra and 
bispectra relevant to the statistics considered in this paper are summarized in Appendix \ref{app:HMformulae}. 
For instance, the covariance of $\epsilon_{0i}$ and $\epsilon_{0j}$ would read
\begin{align}
P_{\epsilon_{0i}\epsilon_{0j}}(k) &= \frac{\delta_{ij}^K}{\bar n_i} 
- b_i \frac{\overline{M_j}}{\bar\rho_m}u(k|M_j) - b_j \frac{\overline{M_i}}{\bar\rho_m} u(k|M_i) 
\nonumber \\
&\qquad + \frac{b_i b_j}{\bar\rho_m^2} \int\!dM\,M^2 n(M) u(k|M)^2 \;.
\label{eq:Pe0e0k}
\end{align}
This expression has the correct limiting (Poissonian) behaviour in the limit $k\to\infty$.
Furthermore, while the Poisson expectation does not depend on $k$ -- which originates from
$\la\tilde\epsilon_{0i}(\vx_1)\tilde\epsilon_{0j}(\vx_2)\ra\sim \delta^D\!(\vx_1-\vx_2)$ -- 
the non-Poissonian corrections involve the low-pass filter $u(k|M)$ since halo exclusion leads to correlations at separations less than
the virial radius (In Lagrangian space, $u(k|M)$ would be replaced by the Lagrangian window function $W(k|M)$, see \cite{Baldauf:2013hka}).
It is beyond the scope of this paper to investigate the scale-dependence of the shot noise any further. 
However, we emphasize that any extension to finite wavenumber is straightforward only if the shot noise is uncorrelated with the density, i.e.
$\la\epsilon_{0}\delta\ra = \la\epsilon_\delta\delta\ra = ... = 0$. 
This may not be the case if halo profiles retain memory of the initial conditions and the halo assembly history. 

Before we discuss the implications of our results from a configuration space point of view, let us emphasize again that,
although Eqs.~(\ref{eq:newHMnoise}) and (\ref{eq:newPTnoise}) is a physically motivated ansatz, there is a priori no reason
why it should furnish better predictions than that of, e.g., \cite{Schmidt:2015gwz}, who derived non-Poissonian expressions
for $B_{\epsilon_{0i}\epsilon_{0j}\epsilon_{0k}}^{\{0\}}$ and $P_{\epsilon_{0i}\epsilon_{\delta j}}^{\{0\}}$ from the conservation of mass and
momentum \cite{1980lssu.book.....P}. This raises the question as to whether there exist model-independent constraints on the
shot noise contributions beyond the usual mass-momentum conservation invoked so far.
We will argue in the forthcoming Section that this is indeed the case.

\section{Shot noise and volume integrals}
\label{sec:configspace}

In this Section, we outline a connection between Poissonian corrections and volume integrals over correlation functions.  
We demonstrate that, whenever certain volume integrals do not vanish, the noise is non-Poissonian. We also establish a correspondence 
between the squeezed limit of the cross halo-matter 3-point functions and the low-$k$ behaviour of the shot noise power spectrum 
$P_{\epsilon_0\epsilon_\delta}^{\{0\}}$, which can be generalized to a new class of consistency, or model-independent relations for biased tracers. 
Remarkably, our new halo model prescription, Eqs.(\ref{eq:newHMnoise}) and (\ref{eq:newPTnoise}), gives results consistent with all these 
expectations.

Before we proceed, let us emphasize that ``volume integral'' refers to the value of the following configuration space integrals:
\begin{equation}
\Xi_{i ... k\delta...\delta} = \int\! d^3r_1 \dots\int\!d^3r_N \, \xi_{i...k\delta...\delta}(\vr_1, ...,\vr_N) \;,
\end{equation}
where the $N$-point correlation function generically involves $k$ halo fields $\delta_1(\vr_1)$, ..., $\delta_k(\vr_k)$ and $N-k$ matter 
fields $\delta(\vr_{k+1})$, ..., $\delta(\vr_N)$. 
Therefore, our discussion does not pertain to the ``integral constraint'' satisfied by correlation function estimators. 

\subsection{2-point correlation function}
\label{sec:2point}

It is standard textbook result that the 2-point correlation of two discrete (homogeneous and isotropic) fluctuation fields 
takes the general form
\begin{equation}
\big\la\delta_i(\vx_1)\delta_j(\vx_2)\big\ra \equiv \frac{1}{\bar n_i}\delta_{ij}^K\delta^D\!(\vr) + \xi_{ij}(r) \;,
\end{equation}
where $r=|{\bf x}_2-{\bf x}_1|$. The first term in the right-hand side is the contribution from self-pairs. 
In the low-$k$ limit, the resulting power spectrum is
\begin{align}
\label{eq:set1}
P_{ij}(k) &= \frac{1}{\bar n_i}\delta_{ij}^K + \int\!d^3r\,\xi_{ij}(r) e^{-i\vk\cdot\vr} \\
&\stackrel{k\ll 1}{=} \left(\frac{1}{\bar n_i}\delta_{ij}^K + \Xi_{ij}\right) + b_i b_j P_\text{lin}(k) 
\nonumber \;,
\end{align}
where
\begin{equation}
\label{eq:defXij}
\Xi_{ij} \equiv \int\!\!d^3 r\,\xi_{ij}(r)
\end{equation}
is generally non-zero even when $i\ne j$ owing to small-scale exclusion effects \cite{Baldauf:2013hka}. 
This implies that cross-power spectra of two different tracer populations have non-zero shot noise contribution, as was 
pointed out in \cite{Hamaus:2010im}. Furthermore, $\Xi_{ii}<0$ (resp. $\Xi_{ii}>0$) leads to sub-Poissonian (resp. 
super-Poissonian) shot noise. 
In \cite{Baldauf:2013hka}, $\Xi_{ij}$ is modelled using peak theory, which exhibits exclusion at small-scales.
Alternatively, we can apply Eq.(\ref{eq:newPTnoise}) to derive a physically consistent prediction for the non-Poissonian 
correction $\Xi_{ij}$.
Namely, equating Eq.(\ref{eq:Pe0e0}) (which follows from Eq.(\ref{eq:newPTnoise}) and Eq.(\ref{eq:set1}) gives
\begin{equation}
P_{\epsilon_{0i}\epsilon_{0j}}^{\{0\}} \equiv \frac{1}{\bar n_i}\delta_{ij}^K + \Xi_{ij} 
\end{equation}
and, consequently,
\begin{equation}
\label{eq:Xij}
\Xi_{ij} \equiv -b_i \frac{\overline{M_j}}{\bar\rho_m}
-b_j \frac{\overline{M_i}}{\bar\rho_m}
+b_i b_j\frac{\la n M^2\ra}{\bar\rho_m^2} \;.
\end{equation}
Note that $\overline{M_i}/\bar\rho_m$ is the Lagrangian volume occupied by the halo. In other words, it is along with the 
average $\la n M^k\ra$ etc., the manifestation of halo exclusion in our halo model framework. 
The last term ensures that a mass weighting of the halo distribution cancels out the noise in the limit where all halos are 
resolved (see Appendix \S\ref{sec:optimal} for a details).
Although only $b_1$ appears, it is pretty clear that higher-order bias has a strong influence on the shape of $\xi_{ij}(r)$,
whence $\Xi_{ij}$. The effect of higher-order bias parameters is, to some extent, captured in the Lagrangian volume 
$\overline{M}/\bar\rho_m$. Finally, note also that, owing to $b_1$, a randomly distributed distribution of hard spheres with 
volume $\sim \overline{M}/\bar\rho_m$ would exhibit a different shot-noise correction than a similar, albeit clustered sample.

\subsection{3-point correlation function}

We now turn to the 3-point correlation function.
With the aid of Eq.(\ref{eq:delta_i}), we arrive at the well-known relations 
\begin{widetext}
\begin{align}
\Big\langle \delta_i(\vx_1) \delta_j(\vx_2) \delta_k(\vx_3) \Big\rangle &=
\xi_{ijk}(\vr_{12},\vr_{13})+\bigg[\frac{\delta_{ij}^K}{\bar n_i}\delta^D\!(\vr_{12})\xi_{ik}(\vr_{13}) + \mbox{(2 cyc.)}\bigg]
+\frac{\delta^K_{ijk}}{\bar n_i^2} \delta^D\!(\vr_{12})\delta^D\!(\vr_{13}) \nonumber \\
\Big\langle \delta_i(\vx_1)\delta_j(\vx_2)\delta(\vx_3)\Big\rangle &=
\xi_{ij\delta}(\vr_{12},\vr_{13}) + \frac{\delta_{ij}^K}{\bar n_i} \delta^D\!(\vr_{12}) \xi_{i\delta}(\vr_{13}) \;.
\end{align}
which, in Fourier space, become
\begin{align}
B_{ijk}(k_1,k_2,k_3) &= \int\! d^3r_{12}\int\!d^3r_{13}\,\xi_{ijk}(\vr_{12},\vr_{13})\, e^{-i \vk_2\cdot\vr_{12}-i\vk_3\cdot\vr_{13}} 
+\bigg[\frac{\delta_{ij}^K}{\bar n_i} \int\! d^3r\,\xi_{jk}(\vr)\, e^{-i\vk_3\cdot\vr} + \mbox{(2 cyc.)}\bigg]
+ \frac{\delta_{ijk}^K}{\bar n_i^2} \\
B_{ij\delta}(k_1,k_2,k_3) &= \int\! d^3r_{12}\int\!d^3r_{13}\,\xi_{ij\delta}(\vr_{12},\vr_{13})\, e^{-i \vk_2\cdot\vr_{12}-i\vk_3\cdot\vr_{13}}
+\frac{\delta_{ij}^K}{\bar n_i}\int\! d^3r\,\xi_{i\delta}(\vr)\, e^{-i \vk_3\cdot\vr} 
\label{eq:BXijdkkk} \;.
\end{align}
\end{widetext}
Hence, in the large scale limit $k_i\to 0$, $B_{ijk}$ and $B_{ij\delta}$ reduce to
\begin{align}
B_{ijk}(k_1,k_2,k_3) &\stackrel{k_i\to 0}{=} \Xi_{ijk} 
+\bigg[\frac{\delta_{ij}^K}{\bar n_i} \Xi_{jk} + \mbox{(2 cyc.)}\bigg]
+ \frac{\delta_{ijk}^K}{\bar n_i^2} \nonumber \\
B_{ij\delta}(k_1,k_2,k_3) &\stackrel{k_i\to 0}{=} \Xi_{ij\delta}+ \frac{\delta_{ij}^K}{\bar n_i} \Xi_{i\delta} \;.
\label{eq:BXijk}
\end{align}
Since the integral of the halo-matter cross-correlations $\xi_{i\delta}(\vr)$ and $\xi_{ij\delta}(\vr_{12},\vr_{13})$, $\Xi_{ij}$ and
$\Xi_{ij\delta}$ respectively, must vanish in order to be consistent with our requirement that $P_{i\delta}(k\to 0)$ and $B_{ij\delta}(k_i\to 0)$ 
be devoid of shot noise, we consistently obtain $B_{\epsilon_{0i}\epsilon_{0j}\delta}^{\{0\}}\equiv 0$. 
In fact, as a general rule, $\Xi_{ij...}$ always vanishes as soon as there is at least one density fluctuation field. 

By contrast, $\Xi_{ijk}$ is not trivial, but our halo model prescription can be used to determine its value. Namely,
\begin{equation}
B_{\epsilon_i\epsilon_j\epsilon_k}^{\{0\}} =
\Xi_{ijk}+\bigg[\frac{\delta_{ij}^K}{\bar n_i}\Xi_{jk} + \mbox{(2 cyc.)}\bigg] + \frac{\delta_{ijk}^K}{\bar n_i^2} 
\label{eq:Bphys}
\end{equation}
upon identifying Eqs. (\ref{eq:bispectra}) and (\ref{eq:BXijk}). Since $\Xi_{ij}$ is already determined by the halo shot noise power 
spectrum $P_{\epsilon_i\epsilon_j}(k)$, we thus obtain a qualitative estimate for the volume integral $\Xi_{ijk}$ of the halo 3-point
function:
\begin{align}
  \label{eq:Xijk}
\Xi_{ijk} &= - b_i b_j b_k \frac{\la n M^3\ra}{\bar\rho_m^3}
+\bigg[ b_i b_j \frac{\overline{M_k^2}}{\bar\rho_m^2} + \mbox{(2 cyc.)}\bigg]  \\ 
&\quad + \bigg[\frac{\delta_{ij}^K}{\bar n_i}b_j \bigg(\frac{\overline{M_k}}{\bar\rho_m}-b_k \frac{\la n M^2\ra}{\bar\rho_m^2}\bigg)
  + \mbox{(2 cyc.)}\bigg]
\nonumber
\end{align}
Our halo model prescription thus provides a simple ansatz for $\Xi_{ijk}$, which could otherwise not be computed unless one is able to 
predict the halo 3-point function $\xi_{ijk}$ deep into the nonlinear regime. Alternatively, clustering statistics of density peaks could 
also be used to furnish a numerical estimate for $\Xi_{ijk}$, and check the cancellation of $\Xi_{ij\delta}$, along the lines of 
\cite{Baldauf:2013hka,Baldauf:2015fbu}.

While it is clear that $B_{XYZ}^{\{0\}}$ should vanish as soon as one of the fluctuation fields is the matter density, it is less obvious 
that $\Xi_{ij\delta}\equiv 0$ for instance. Namely, one might expect that, since two halo fields are present, small-scale exclusion effects
may conspire and lead to $\Xi_{ij\delta}\ne 0$. The relation between the shot noise contributions in Fourier space and volume integrals 
in configuration space clearly shows that this cannot be true. This will be confirmed by our measurements of $B_{ii\delta}$, which do not 
exhibit any significant constant white noise for $k_i\ll 1$ (although they admittedly do not rule out a small value of $\Xi_{ij\delta}$).

\subsection{Shot noise and squeezed limit}
\label{sec:squeezed}

So far, we have demonstrated how $P_{\epsilon_0\epsilon_0}^{\{0\}}$ and $B_{\epsilon_0\epsilon_0\epsilon_0}^{\{0\}}$ are related to $\Xi_{ij}$ and
$\Xi_{ijk}$. Can we find a similar correspondence for $P_{\epsilon_0\epsilon_\delta}^{\{0\}}$ ? To answer this question, let us scrutinize 
Eq.(\ref{eq:BXijdkkk}). For small albeit non-zero $k_3$, the second term in the right-hand side is
\begin{equation}
\frac{\delta_{ij}^K}{\bar n_i} \int\!d^3r\,\xi_{i\delta}(\vr)\,e^{-i\vk_3\cdot\vr} \approx \frac{b_i}{\bar n_i}P_\text{lin}(k_3) \delta_{ij}^K
\end{equation}
since $\Xi_{i\delta}\equiv 0$. This is precisely the Poissonian expectation of 
$P_{\epsilon_{0i}\epsilon_{\delta j}}^{\{0\}}+P_{\epsilon_{0j}\epsilon_{\delta i}}^{\{0\}}$. The non-Poissonian correction is thus hiding in the
other term. To extract the contribution proportional to $P_\text{lin}(k_3)$, we consider triangular configurations with $k_i\ll 1$,
so that the first term in the right-hand side of Eq.(\ref{eq:BXijdkkk}) is approximately
\begin{equation}
\label{eq:xijdL1}
\int\!d^3r_{13}\bigg\{\int\! d^3r_{12}\,\xi_{ij\delta}(\vr_{12},\vr_{13})\bigg\} e^{-i\vk_3\cdot\vr_{13}}\;.
\end{equation}
Since, by convention, $\vk_3$ is the wavevector carried by the density fluctuation, the 3-point correlation $\xi_{ij\delta}(\vr_{12},\vr_{13})$ 
is computed in the limit of a ``soft'' or long-wavelength density fluctuation $\delta(\vk_3)\equiv \delta_\text{L}$:
\begin{align}
\label{eq:dxiddL}
\xi_{ij\delta}(\vr_{12},\vr_{13}) &\stackrel{k_3\ll 1}{=}\big\la \xi_{ij}\big(\vr_{12}|\delta_\text{L}\big)\delta_\text{L}\big\ra \\
&~\approx \frac{\partial}{\partial\delta_\text{L}}\xi_{ij}(\vr_{12}|\delta_\text{L})\bigg\lvert_{\delta_\text{L}=0}
\la \delta_\text{L}^2\ra \nonumber \;,
\end{align}
which follows from Bayes' theorem and from a Taylor-expansion of $\xi_{ij}(\vr_{12}|\delta_\text{L})$ at first-order.
Ref.\cite{Sherwin:2012nh} computed the derivative $\partial\xi_{ij}/\partial\delta_\text{L}$ in the limit of large separations 
$|\vr_{12}|\gg R_M$, where $R_M$ is the characteristic halo scale 
(see also, e.g., \cite{Baldauf:2011bh,Kehagias:2013yd} for the Fourier space calculation).
However, our task is fundamentally different here mainly for two reasons: We need to calculate the derivative w.r.t. the long mode 
$\delta_\text{L}$ when $\xi_{ij}$ is integrated over all $\vr_{12}$ and, most importantly, extract the response of the shot noise or, 
equivalently, the piece uncorrelated with the density fluctuation field. 
To explain this seemingly inconsistent statement, recall that the fluctuation fields in Eq.(\ref{eq:dhPT}) include all perturbations but
the long mode. In other words, the condition $\la\epsilon_0\delta\ra = \la\epsilon_\delta\delta\ra = ... = 0$ and the dependence of the 
shot noise on $\delta_\text{L}$ are not mutually exclusive because the long mode can be absorbed into a local rescaling of the cosmology,
which the shot noise should be sensitive to.

Substituting Eq.(\ref{eq:dxiddL}) into Eq.(\ref{eq:xijdL1}), we obtain
\begin{align}
\label{eq:eq:xijdL2}
\int\!d^3r_{13}\bigg\{\int\! &d^3r_{12}\,\xi_{ij\delta}(\vr_{12},\vr_{13})\bigg\} e^{-i\vk_3\cdot\vr_{13}} \\
&\approx \bigg\{\int\! d^3r_{12}\,\frac{\partial \xi_{ij}}{\partial\delta_\text{L}}(\vr_{12}|\delta_\text{L})\bigg\lvert_{\delta_\text{L}=0}
\bigg\}P_\text{lin}(k_3) \nonumber \\
&=\frac{\partial}{\partial\delta_\text{L}}\bigg\{\int\! d^3r_{12}\,\xi_{ij}(\vr_{12}|\delta_\text{L})\bigg\}_{\delta_{L}=0}P_\text{lin}(k_3) 
\nonumber \\
&= \bar\rho_m \frac{\partial \Xi_{ij}}{\partial\tilde\rho_m}\bigg\lvert_{\tilde\rho_m=\bar\rho_m} P_\text{lin}(k_3) \nonumber \;.
\end{align}
In the last equality, we have substituted the definition of $\Xi_{ij}$ as volume integral over $\xi_{ij}$. 
Moreover, the derivative is taken w.r.t. the average matter density $\tilde\rho_m$, and subsequently evaluated at $\tilde\rho_m=\bar\rho_m$. 
For notational purposes, we shall hereafter denote this derivative by $\partial\Xi_{ij}/\partial\bar\rho_m$. 
Therefore, the limit $k_i\ll 1$ of Eq.(\ref{eq:BXijdkkk}) gives the consistency (model-independent) relation
\begin{equation}
P_{\epsilon_{0i}\epsilon_{\delta j}}^{\{0\}}+P_{\epsilon_{0j}\epsilon_{\delta i}}^{\{0\}} 
= \frac{b_i}{\bar n_i}\delta_{ij}^K + \bar\rho_m\frac{\partial\Xi_{ij}}{\partial\bar\rho_m}
\label{eq:noiseconsistency}
\end{equation}
which must be satisfied by any realistic, discrete tracer of the large scale structure.
This illustrates how shot noise corrections can be computed from the squeezed or soft limit of configuration space integrals. 
Eq.(\ref{eq:noiseconsistency}), together with its obvious generalization to higher-order correlators, provide a new class of 
(model-independent) consistency relations for biased tracers which have not been considered yet in the literature. 

Let us now assess whether our halo model prescription satisfies Eq.(\ref{eq:noiseconsistency}).
To proceed further, we must bear in mind that our task is to extract the response of the halo shot noise to the long mode. Therefore, we 
shall not take derivatives of, e.g. $b_i$ and $b_j$ in Eq.(\ref{eq:Xij}) because they would describe contributions correlated with the 
density field $\delta(\vx)$. Derivatives of the $\bar\rho_m$ factors should also be discarded on similar grounds. 
Since we already emphasized that non-Poissonian corrections are closely related to halo exclusion, it should be clear that the halo 
exclusion volume or, since $V_\text{exc}\sim M/\bar\rho_m$ and $\bar\rho_m$ is unchanged, the halo mass $M$ is the quantity we should 
compute the response of. Therefore, the derivative $\partial\Xi_{ij}/\partial\bar\rho_m$ is given by
\begin{equation}
\frac{\partial\Xi_{ij}}{\partial\bar\rho_m} = 
- \frac{b_i}{\bar\rho_m}\frac{\partial\overline{M_j}}{\partial\bar\rho_m}
- \frac{b_j}{\bar\rho_m}\frac{\partial\overline{M_i}}{\partial\bar\rho_m}
+\frac{b_i b_j}{\bar\rho_m^2} \frac{\partial\la n M^2\ra}{\partial\bar\rho_m}\;.
\label{eq:dXijddL}
\end{equation}
To evaluate the derivatives of $\overline{M}$ and $\la n M^2\ra$ w.r.t. $\bar\rho_m$, we hold the halo mass scale fixed, but vary the 
halo number density that appears in these two averages. This would effectively corresponds to a change in exclusion volume (which is
certainly proportional to the abundance of halos in a given region of space). We find:
\begin{align}
\frac{\partial \overline{M_i}}{\partial\bar\rho_m} &= \frac{\partial}{\partial\bar\rho_m}\left(\frac{1}{\bar n_i}\int\!dM\,M n\, 
\Theta(M,M_i)\right) \nonumber \\
&= \frac{1}{\bar n_i} \int\!dM\,M \left(\frac{\partial n}{\partial\bar\rho_m}\right)\Theta(M,M_i) \nonumber  \\
&= \frac{1}{\bar n_i\bar\rho_m}\int\!dM\,M n \left(\frac{n}{\bar\rho_m}\frac{\partial n}{\partial\bar\rho_m}\right)\Theta(M,M_i) 
\nonumber \\
&= \frac{\overline{M_ib_i}}{\bar\rho_m} \;.
\end{align}
Clearly, $M>M_\star$ halos with $b_1 > 1$ (here, $M_\star=M_\star(z)$ is the characteristic mass of halos collapsing at redshift $z$) have
their small-scale exclusion amplified in the presence of a long mode because their number density locally increases, so that there is 
less volume available to new halos. Similarly, 
\begin{align}
\frac{\partial\la n M^2\ra}{\partial\bar\rho_m}
&= \int\!dM\,M^2 \left(\frac{\partial n}{\partial\bar\rho_m}\right) \nonumber \\
&= \frac{\la n M^2 b_1\ra}{\bar\rho_m} \;.
\end{align}
Substituting these expressions into Eq.(\ref{eq:dXijddL}), the derivative of $\Xi_{ij}$ w.r.t. the long mode can be recast into the
form
\begin{equation}
\bar\rho_m\frac{\partial\Xi_{ij}}{\partial\bar\rho_m} = 
- b_i\frac{\overline{M_jb_j}}{\bar\rho_m}
- b_j\frac{\overline{M_ib_i}}{\bar\rho_m}
+b_i b_j \frac{\la n M^2 b_1\ra}{\bar\rho_m^2} \;,
\end{equation}
which precisely agrees with the non-Poissonian correction in Eq.(\ref{eq:Pe0ed}).
Remarkably, our new halo model prescription based on Eqs.(\ref{eq:newHMnoise}) and (\ref{eq:newPTnoise}) satisfies the consistency 
relation Eq.(\ref{eq:noiseconsistency}). 

\vspace{5mm}

To conclude this Section, our halo model approach straightforwardly generalizes to higher-order noise covariances, and the correspondence
between between Fourier space shot noise covariances and configuration space integrals can be used to derive quantitative estimates for
$\Xi_{ijkl...}$ and their derivatives w.r.t. to the background density $\bar\rho_m$. 
For instance, the covariance of band-power average halo power spectrum $\hat P_{hh}(k)$,
\begin{equation}
C_{ij}\equiv \big\la \hat P_{hh}(k_i)\hat P_{hh}(k_j)\big\ra-\big\la\hat P_{hh}(k_i)\big\ra \big\la\hat P_{hh}(k_j)\big\ra \;,
\end{equation}
is the sum of a Gaussian contribution proportional to the square of the halo power spectrum, and a non-Gaussian piece proportional 
to the halo trispectrum \cite{Scoccimarro:1999kp}.
A similar relation holds for the covariance of the cross halo-matter power spectrum.
Assuming Poisson noise, Ref.\cite{Smith:2008ut} and \cite{Chan:2016ehg} derive these two contributions to the covariance of 
the cross- and auto-power spectrum, respectively. 
However, in light of our measurements of the halo noise bispectrum, we expect similar corrections to the noise trispectrum etc. which
would propagate into the covariance of band-power measurements. 
The interested reader can find in Appendix \S\ref{app:trispec} the calculation of the halo noise 4-point function in the limit 
$k_i\to 0$, along with our halo model prediction for the trispectrum white noise 
$T_{\epsilon_{0i}\epsilon_{0j}\epsilon_{0k}\epsilon_{0l}}^{\{0\}}$.

\section{Comparison to N-body simulations}
\label{sec:nbody}

In this Section, we assess the accuracy of our new halo model ansatz summarized by Eqs. (\ref{eq:newHMnoise}) and (\ref{eq:newPTnoise}).
Namely, we test our predictions for $P_{\epsilon_{0i}\epsilon_{\delta j}}^{\{0\}}$ (Eq.(\ref{eq:Pe0ed})) and 
$B_{\epsilon_{0i}\epsilon_{0j}\epsilon_{0k}}^{\{0\}}$ (Eq.(\ref{eq:Be0e0e0})) using measurements of cross halo-matter bispectra extracted from 
numerical simulations. We focus on shot noise contributions for the same halo catalogue (i.e., $i=j=k$ in the relevant formulae). 

\subsection{Numerical simulations}

We use a large suite of 512 N-body simulations from the DEUS-PUR project  \cite{Chan:2016ehg,Blot:2014pga,Rasera:2013xfa}. 
The cosmology is a $\Lambda$CDM model consistent with the {\small WMAP7} constraints \cite{Spergel:2006hy}: a Hubble parameter $h=0.72$, matter
density $\Omega_m = 0.257$, scalar index $n_s=0.963$ and normalization amplitude $\sigma_8=0.801$. 
Each simulation evolves $512^3$ particles in a cubical box of size $1312.5\hmpc$. The mass of a dark matter particle thus is 
$1.2\times 10^{12}\hmsun$.
Gaussian initial conditions were laid down with the Zel'dovich approximation at the initial redshift $z_i=105$, and subsequently evolved using 
the {\small RAMSES} solver \cite{Teyssier:2001cp}.
Snapshots of the dark matter distribution were produced at redshift $z=0$, 0.5 and 1. They were post-processed with a FoF algorithm of linking
length 0.2 times the mean inter-particle separation in order to produce the halo catalogues analyzed here.

We discard halos with less than 100 particles, which yields a minimum halo mass $M_\text{min}=1.20 \times 10^{14}\hmsun$.
Furthermore, since the number density of these halos is already very low, we do not divide them further into mass bins. Hence, we have only
one linear bias parameter $b_1$ per snapshot, which characterizes the clustering of all $M\geq M_\text{min}$ halos. 
To estimate $b_1$ for the three different halo samples, we measure the power spectrum ratio \cite{Hamaus:2010im}
\begin{equation}
\label{eq:b_idef}
\hat{b}_1=\frac{P_{h\delta}(k)}{P_{\delta\delta}(k)}
\end{equation}
as it is unaffected by shot noise.

We calculate $b_1$ separately for each realization before averaging over the whole set of realizations. To estimate $b_1$, we consider all the
Fourier modes with $k<\kmax$. We choose $\kmax = 0.02 h \mathrm{Mpc^{-1}}$, which is a good compromise between a smaller $\kmax$ 
leading to a larger sampling variance, and a larger $\kmax$ with wavemodes potentially affected by nonlinearities. Our best-fit values of
$b_1$, along with the number density and average mass of each halo catalogue, are summarized in Table \ref{table1}.

We note that, although we use the same 'Large set' of simulations as \cite{Chan:2016ehg}, our linear bias estimates are slightly different 
because \cite{Chan:2016ehg} computed $b_1$ using the halo-halo power spectrum (which is affected by white noise in the limit $k\to 0$).

\begin{table}
\caption{Redshift $z$, linear bias $b_1$, number density $\bar n$ (in $\hhhmpc$) and average mass $\overline{M}$ (in $\hmsun$) of the three 
halo catalogues harvested for the comparison with the theoretical predictions.}
\vspace{1mm}
\begin{center}
\begin{tabular}{c|c|c|c}
\hline
$z$ & $b_1$ & $\bar n$ & $\overline{M}$ \\
\hline\hline
~~~z=0~~~ & ~~~3.08~~~  & ~~~1.57$\times 10^{-5}$~~~ & ~~~2.59$\times 10^{14}$~~~  \\
\hline
z=0.5 & 4.47 & 6.70$\times 10^{-6}$ & 2.13$\times 10^{14}$ \\
\hline
z=1 & 6.46 & 1.92$\times 10^{-6}$ & 1.84$\times 10^{14}$ \\
\hline
\end{tabular}
\end{center}
\label{table1}
\end{table}

\subsection{Power spectra and Bispectra}

Since the bispectrum is significantly noisier than the power spectrum, we calculate instead the average cumulative bispectrum as a function
of a maximum wavenumber $\kmax$, which we define as
\begin{equation}
\label{eq:defbmax}
B_{XYZ}(<\kmax)\equiv\frac{V^2}{N_t}\sum X(\vk_1) Y(\vk_2) Z(\vk_3)\;.
\end{equation}
Here, $X(\vk_1)$, $Y(\vk_2)$ and $Z(\vk_3)$ are the Fourier transform of the gridded (discrete) fluctuation fields corrected for the window
assignement function.
The sum runs over all wavevectors $\vk_i$ satisfying the condition 
$0< k_i <\kmax$ and the momentum conservation $\vk_1+\vk_2+\vk_3=0$. $N_t$ is the total number of such admissible triangular configurations. 
Note that the reality condition for, e.g.,  the density field $\delta$ implies $\delta(\vk)= \delta^*(-\vk)$.
In this case, the sum can also be written as 
\begin{equation}
\label{eq:defbmax2}
B_{\delta\delta\delta}(<\kmax)=\frac{V^2}{N_t}\sum \delta(\vk_1) \delta(\vk_2) \delta^*(\vk_1+\vk_2)\;
\end{equation}
where the sum runs over all the pairs ${\vk_1,\vk_2}$ of wavevectors such that $\{0<k_1,k_2,|\vk_1+\vk_2|<\kmax\}$ is satisfied. 

\begin{figure*}
\subfloat{\includegraphics[width=0.42\textwidth]{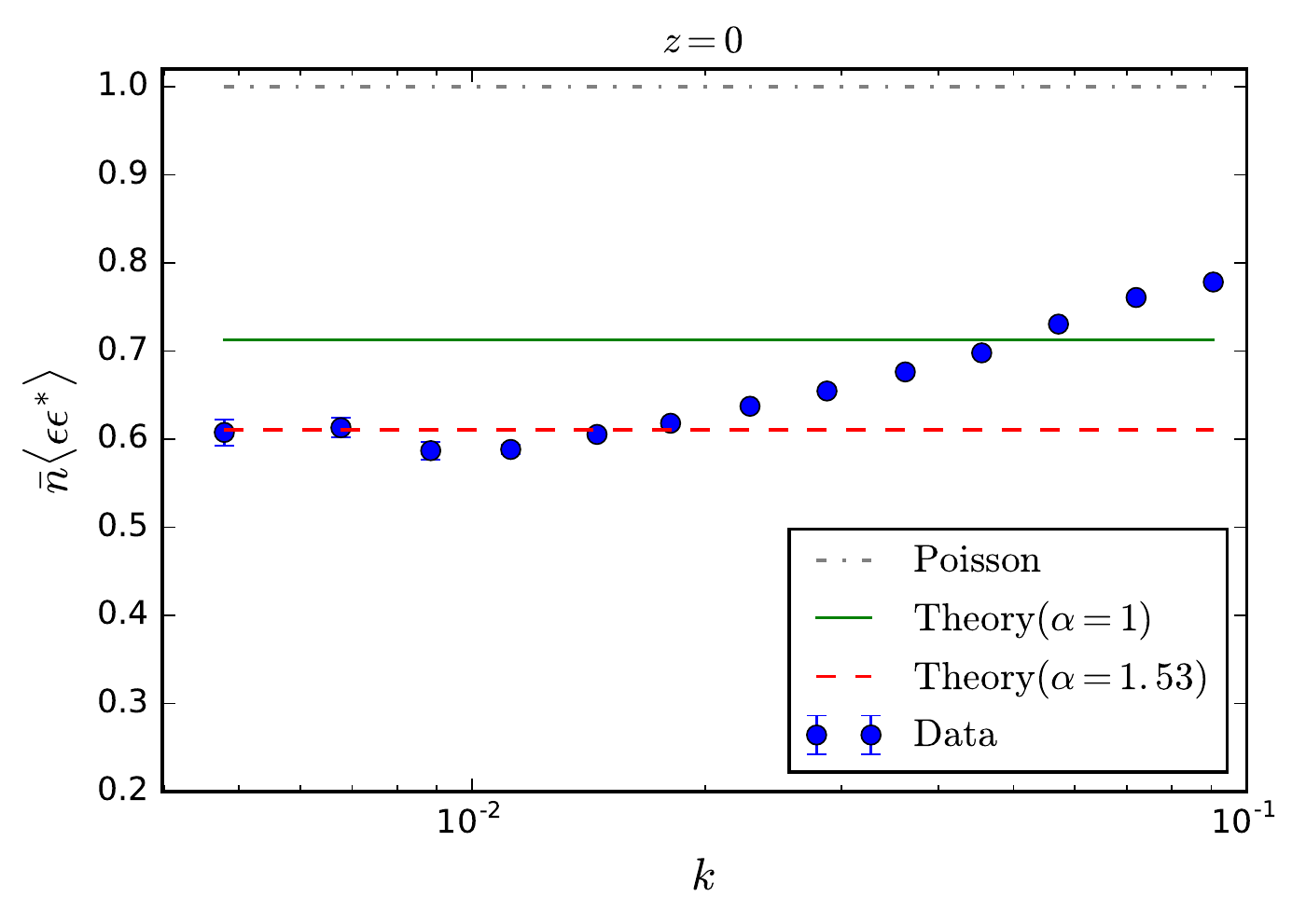}}
\subfloat{\includegraphics[width=0.42\textwidth]{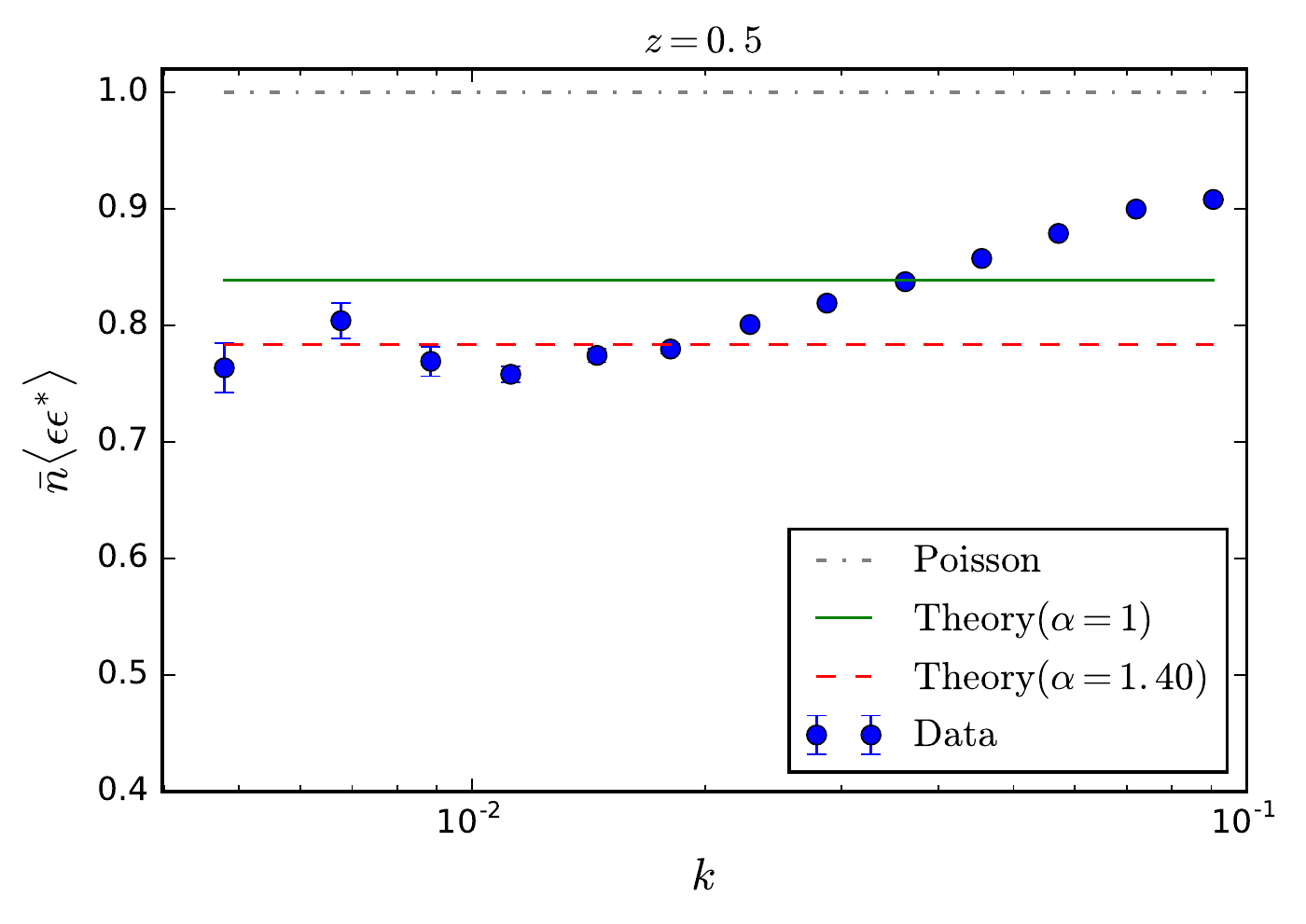}} \\
\subfloat{\includegraphics[width=0.42\textwidth]{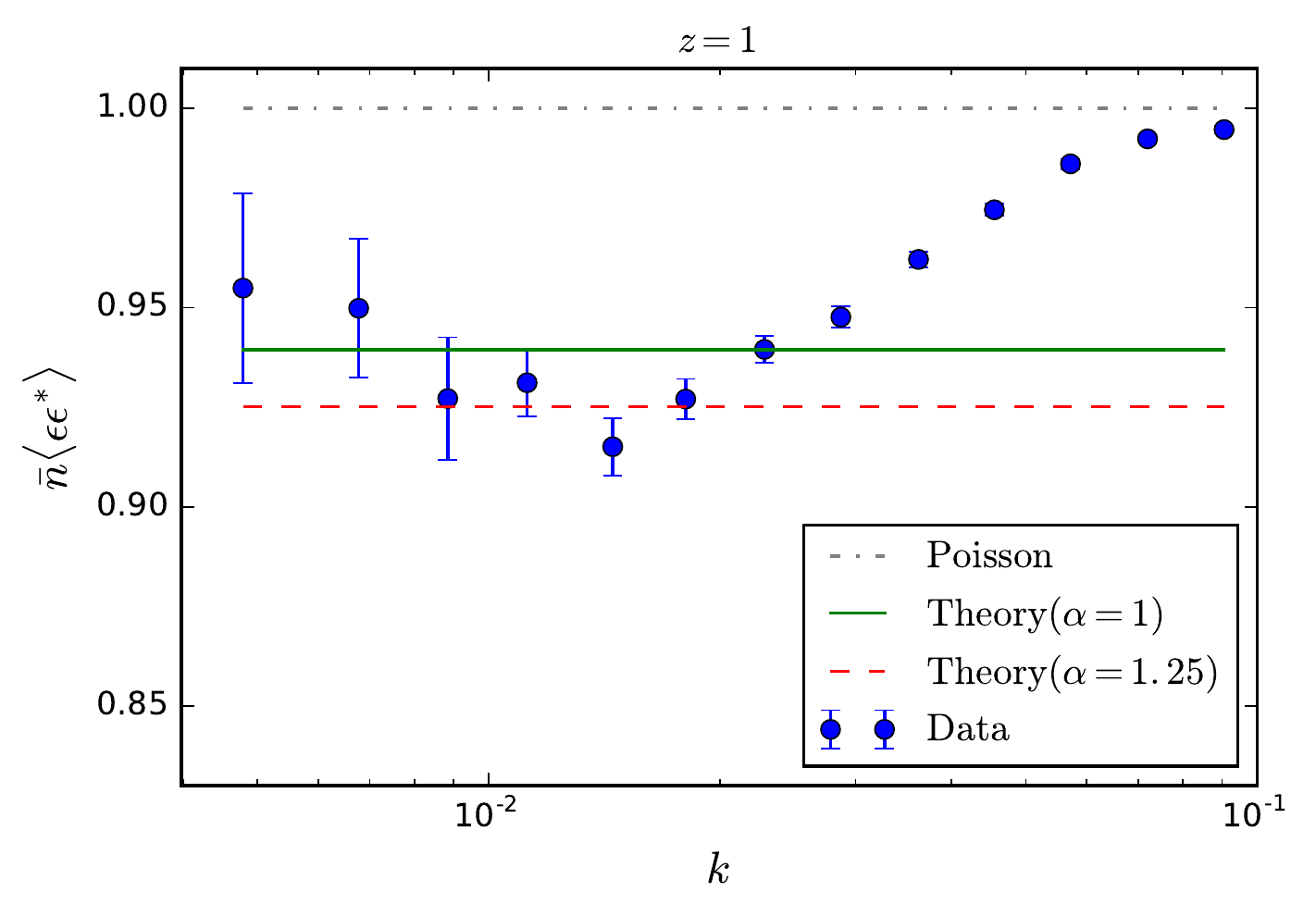}}
\caption{Noise power spectrum $P_{\epsilon\epsilon}(k)$ normalized to the Poisson expectation $1/\bar n$ as a function of wavenumber 
Results are shown for three different catalogues of massive halos, $M\gtrsim 10^{14}\hmsun$, spanning the redshift range $0<z<1$. 
The dotted-dashed (grey) indicates the Poisson expectation, while the solid (green) and dashed (red) lines represent the halo model 
prediction $P_{\epsilon_0\epsilon_0}^{\{0\}}$ (see Eq.(\ref{eq:eqswithalpha}) with $\alpha=1$ and $\alpha$ fitted to the low-$k$ data points, 
respectively.}
\label{fig:Pee}
\end{figure*}

In order to test our algorithm, we generated 512 realizations of randomly distributed particles with the same number density as that of the
halos extracted from the simulations. To compute the Fourier modes, we extrapolated these random distributions onto a regular cubical grid
using the cloud-in-cell (CiC) assignment which, upon Fourier transformation, was corrected for with the appropriate window function. The
resulting cumulative bispectrum was consistent with $1/\bar n^2$ within the error bars. 
A comparison of the cumulative matter bispectrum with the tree-level expression Eq.(\ref{eq:bddd}) induced by gravitational mode-coupling 
provided a second consistency check. We found a very good agreement between the simulated and theoretical $B_{\delta\delta\delta}(<\kmax)$, 
with a systematic deviation towards $\kmax\sim 0.1\hmmpc$ presumably consistent with the gradual rise of 1-loop contributions to the matter
bispectrum. 

Hereafter, we will discuss results obtained for the noise power spectrum, and  the following cumulative bispectrum of halo noise and matter 
fluctuation fields:
\begin{align}
\label{eq:Bmax2}
&B_{\epsilon\epsilon\epsilon}(<\kmax)=\frac{V^2}{N_t}\sum \epsilon(\vk_1) \epsilon(\vk_2) \epsilon^*(\vk_1+\vk_2)\nonumber\\ 
&B_{\epsilon\epsilon\delta}(<\kmax)=\frac{V^2}{N_t}\sum \epsilon(\vk_1) \epsilon(\vk_2) \delta^*(\vk_1+\vk_2)\;,
\end{align}
where the sum runs, again, over all pairs $\{\vk_1,\vk_2\}$ such that ${0<k_1,k_2,|\vk_1+\vk_2|<\kmax}$. The halo noise $\epsilon(\vk)$ is 
defined as in Eq.(\ref{eq:defeps}), $\epsilon(\vk)=\delta_h(\vk)-b_1\delta(\vk)$, where $b_1$ assumes the best-fit values listed in Table
\ref{table1}. Since it is pausible that the exclusion volume responsible for sub-Poissonian noise at large halo mass be different from  
$\overline{M}/\bar\rho_m$ (after all, halos do not have well defined boundaries, but rather a density profile which decays with the distance 
from the halo center), we multiply in Eqs.(\ref{eq:Pe0e0ii}), (\ref{eq:Pe0edii}) and (\ref{eq:Be0e0e0iii}) each factor of $M$ by a fudge 
factor $\alpha$. With this additional model parameter, our theoretical predictions based on Eqs.(\ref{eq:newHMnoise} and (\ref{eq:newPTnoise})) 
thus become:
\begin{align}
P_{\epsilon_0\epsilon_0}^{\{0\}} &= \frac{1}{\bar n}\bigg\{1 -2\alpha b_1\left(\frac{\bar n\overline{M}}{\bar\rho_m}\right)
+\alpha^2b_1^2\left(\frac{\bar n \la n M^2\ra}{\bar\rho_m^2}\right)\bigg\} \nonumber \\
P_{\epsilon_0\epsilon_\delta}^{\{0\}} &= \frac{b_1}{2\bar n}\bigg\{1 -2\alpha\left(\frac{\bar n\overline{Mb_1}}{\bar\rho_m}\right)
+\alpha^2b_1\left(\frac{\bar n \la n M^2 b_1\ra}{\bar\rho_m^2}\right)\bigg\} \nonumber \\
B_{\epsilon_0\epsilon_0\epsilon_0}^{\{0\}} &=\frac{1}{\bar n^2}
\bigg\{1 - 3\alpha b_1\left(\frac{\bar n \overline{M}}{\bar\rho_m}\right) +3\alpha^2b_1^2 \left(\frac{\bar n^2 \overline{M^2}}{\bar\rho_m^2}
\right) \nonumber \\
&\qquad - \alpha^3b_1^3 \left(\frac{\bar n^2\la n M^3\ra}{\bar \rho_m^3}\right) \bigg\} \;.
\label{eq:eqswithalpha}
\end{align}
Note that, while $P_{\epsilon_0\epsilon_0}^{\{0\}}$ is nothing but the formula derived in \cite{Hamaus:2010im}, the other expressions are 
new. To evaluate the average $\la n M^k b_1^l\ra$, we take the lower boundary of the integral to be equal to the mass of one dark 
matter particle, i.e. $1.2\times 10^{12}\hmsun$. We use the Sheth-Tormen mass function and bias for this numerical evaluation 
\cite{Sheth:1999mn}, which matches the measured halo number densities to better than 10 percents at all redshifts.
We have found that $\la n M^k b_1^l\ra$ does not change noticeably if we decrease further the lower bound of the integral. 
Furthermore, we assume $\overline{Mb_1}\approx \overline{M} b_1$ to rely as much as possible on the measured mass, number density and
bias for the halo bin. The Sheth-Tormen mass function and bias suggest that $\overline{M} b_1$ underestimates the actual average 
$\overline{Mb_1}$ by $\sim 15$\% (resp. $\sim$5 \%) at $z=0$ (resp. $z=1$).

\begin{figure*}
\subfloat{\includegraphics[width=0.42\textwidth]{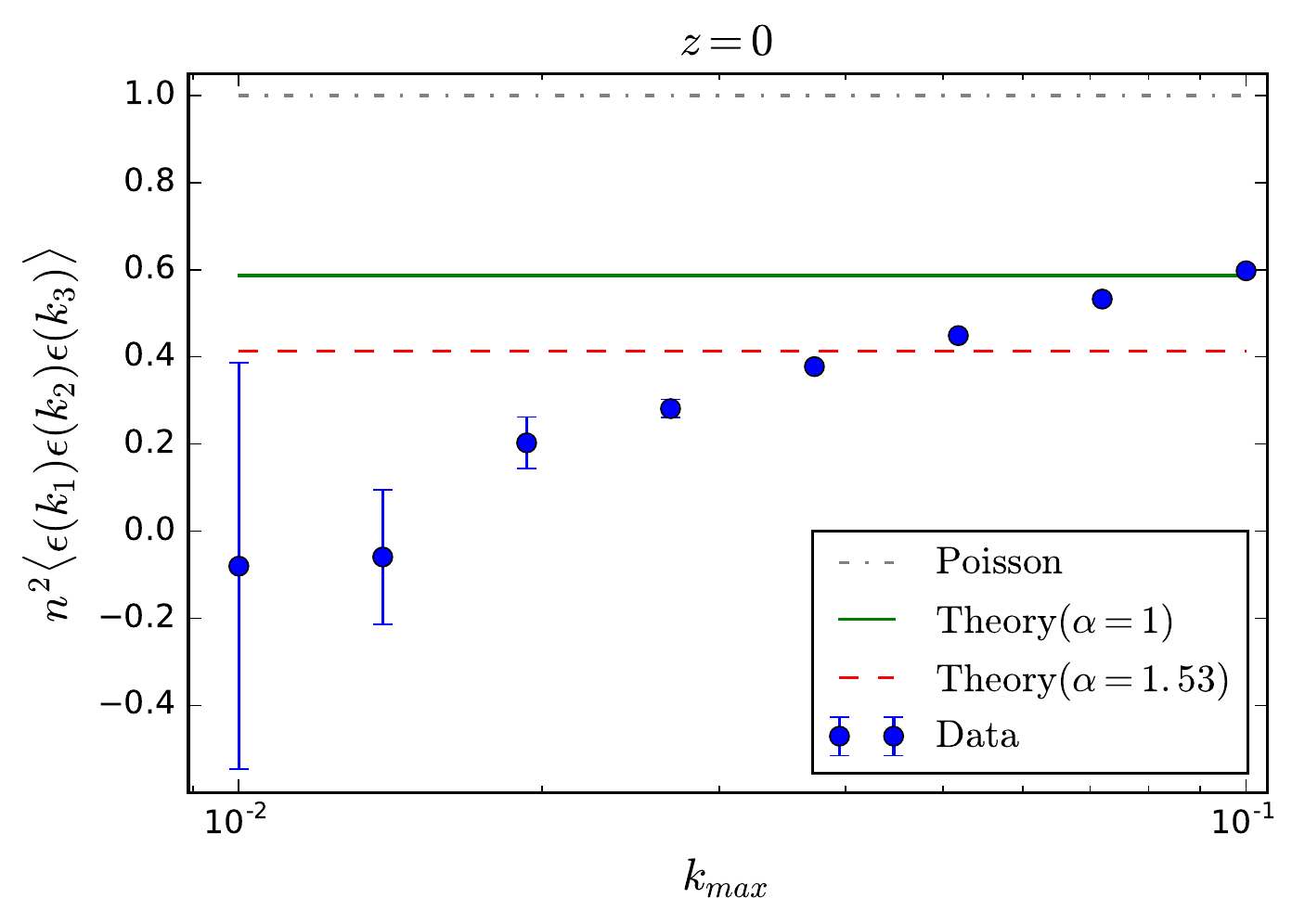}}
\subfloat{\includegraphics[width=0.42\textwidth]{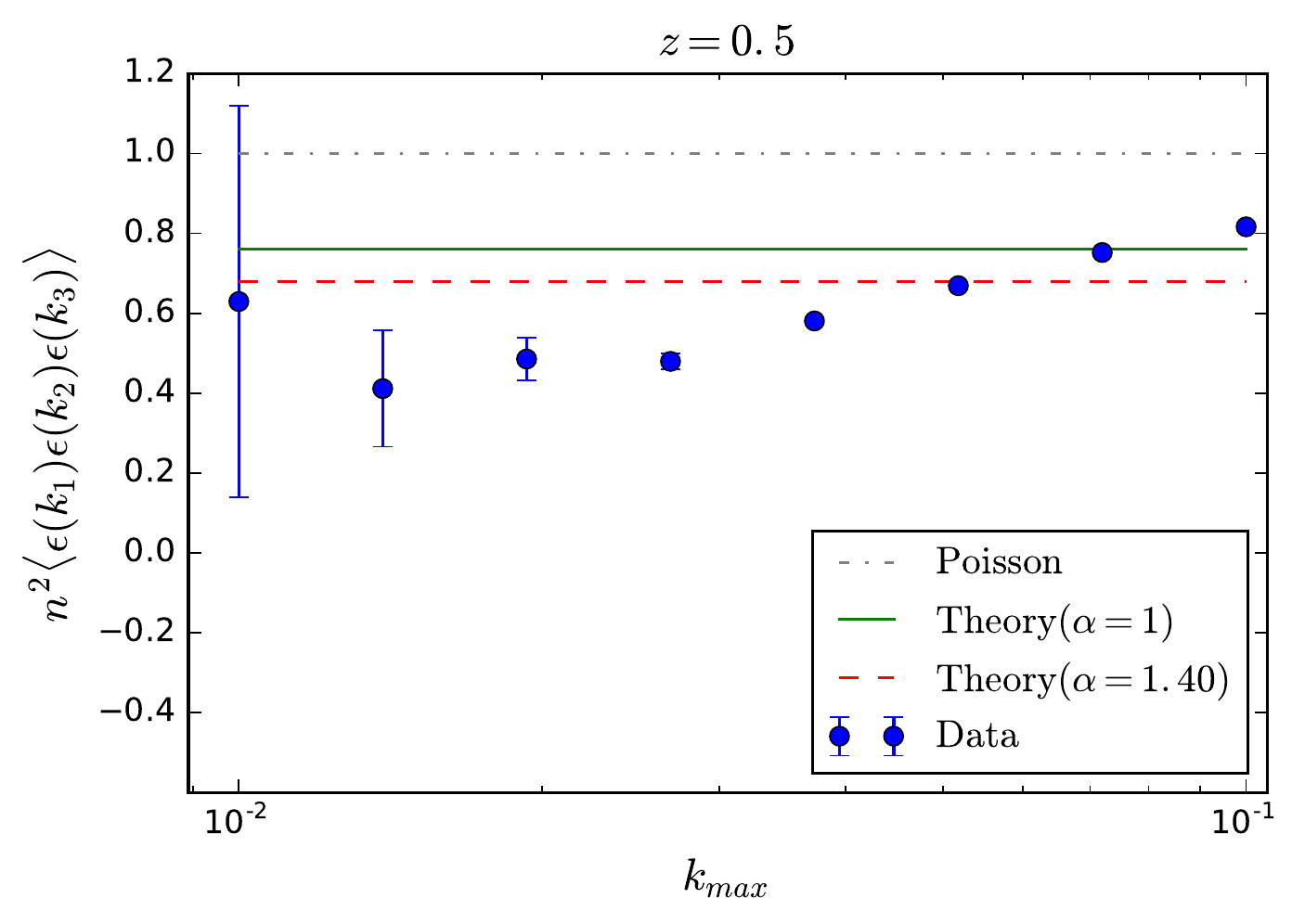}} \\
\subfloat{\includegraphics[width=0.42\textwidth]{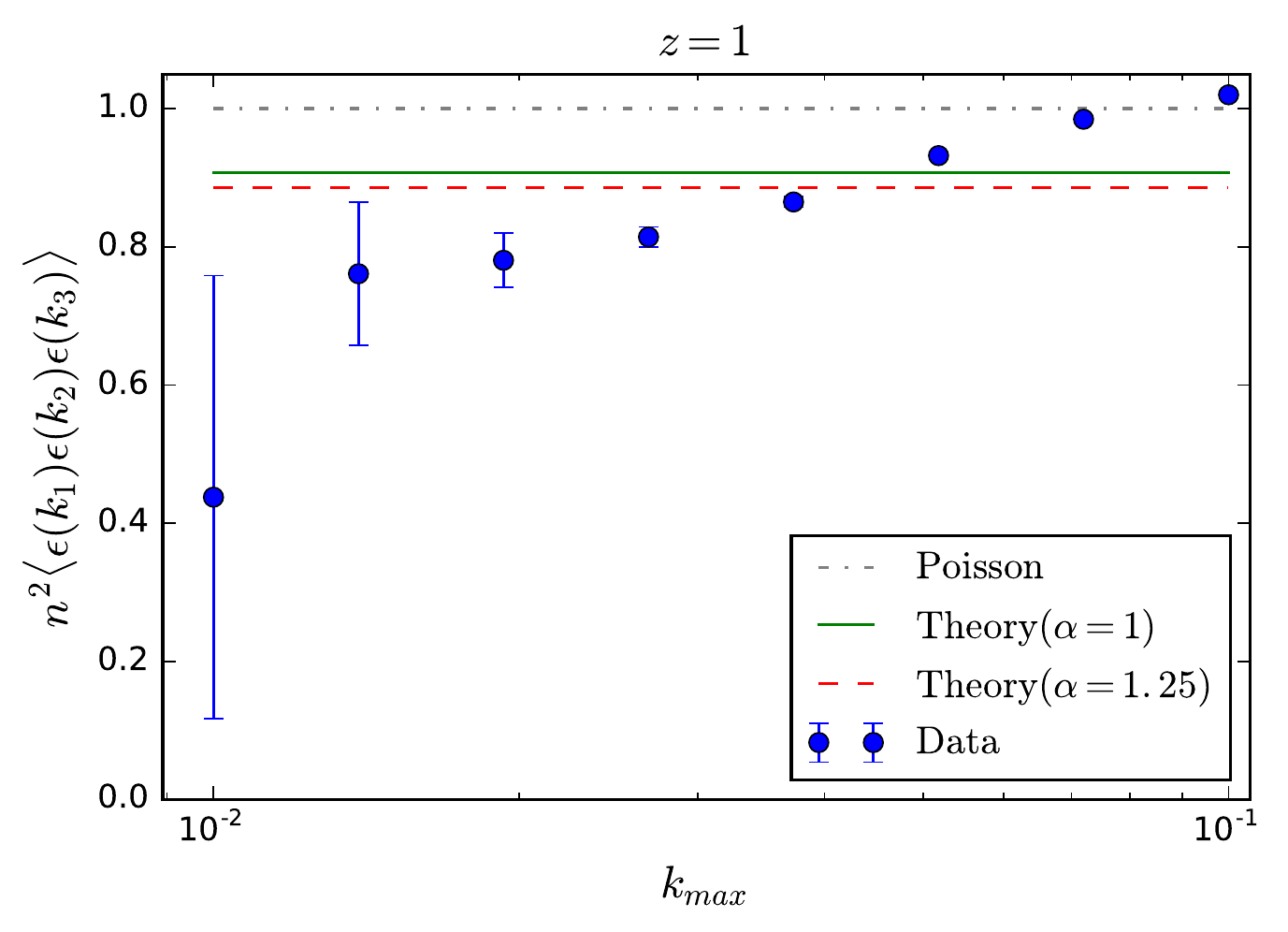}}
\caption{Measurements of the cummulative halo noise bispectrum $B_{\epsilon\epsilon\epsilon}(<\kmax)$ normalized to the Poisson
expectation $1/\bar n^2$ as a function of the maximum wavenumber $\kmax$. Results are shown at redshift $z=0$, 0.5 and 1. 
Error bars indicate the error on the mean as estimated from the full suite of simulations. The solid (green) lines is the halo
model prediction with $\alpha=1$, whereas the dashed (red) line follows from setting $\alpha$ to its best-fit value inferred 
from the measurements of the halo noise power spectrum.}
\label{fig:Beee}
\end{figure*}

\subsection{Results}

We first calculate the noise power spectrum $P_{\epsilon\epsilon}(k)$ and find that it lies significantly below the Poisson expectation 
$1/\bar n$, in agreement with the findings of \cite{Hamaus:2010im}. 
This is summarized in Fig.\ref{fig:Pee}. 
The error bars represent the standard deviation of the mean, which is calculated from 512 realizations.
The solid (green) curve indicates $P_{\epsilon_0\epsilon_0}^{\{0\}}$ for the fiducial value of $\alpha=1$, whereas the dashed (red) curve
is a fit to the low-$k$ plateau seen in the figure (For $z=1$, we discarded the two data points with lowest wavenumbers). 
The best-fit values of $\alpha$ are quoted in the insert window. We will use the same values of $\alpha$ in subsequent comparison in
order to assess the extent to which a single value of $\alpha=\alpha(M)$ can reproduce all the measurements.
Furthermore, $P_{\epsilon\epsilon}(k)$ exhibits a strong scale-dependence for $k\gtrsim 0.2\hmmpc$ similar to that seen in 
\cite{Biagetti:2016ywx}. If the $k$-dependence of the noise power spectrum directly reflects that of the halo profile $u(k|M)$, as 
in Eq.(\ref{eq:Pe0e0k}) for instance, then it cannot explain the scale-dependence seen in Fig.\ref{fig:Pee} because $u(k|M)$ 
remains very close to unity for $k\lesssim 0.1\hmmpc$, even for massive halos. Clearly, the observed scale-dependence must therefore 
originate from the second- and higher- order terms in the definition of the halo shot noise, Eq.(\ref{eq:defeps}).  

We now turn to the measurements of the cumulative halo noise bispectrum $B_{\epsilon\epsilon\epsilon}(<\kmax)$. 
They are shown in Fig.\ref{fig:Beee} as a function of the maximum wavenumber $\kmax$. 
Here again, the data lies consistently below the Poisson expectation $1/\bar n^2$. 
Unfortunately, the lowest data point is fairly noisy, so that it is difficult to identify any plateau which could unambigously help 
us determine the actual magnitude of $B_{\epsilon_0\epsilon_0\epsilon_0}^{\{0\}}$. The scale-dependence, which is more pronounced than in  
$P_{\epsilon\epsilon}(k)$, complicates this further. Moreover, using the best-fit value of $\alpha$ inferred from the measurements of 
$P_{\epsilon\epsilon}(k)$ only mildly improve the agreement at low wavenumber. At $z=0$ for instance, matching the data would require
$\alpha \approx 2$.

\begin{figure*}
\subfloat{\includegraphics[width=0.42\textwidth]{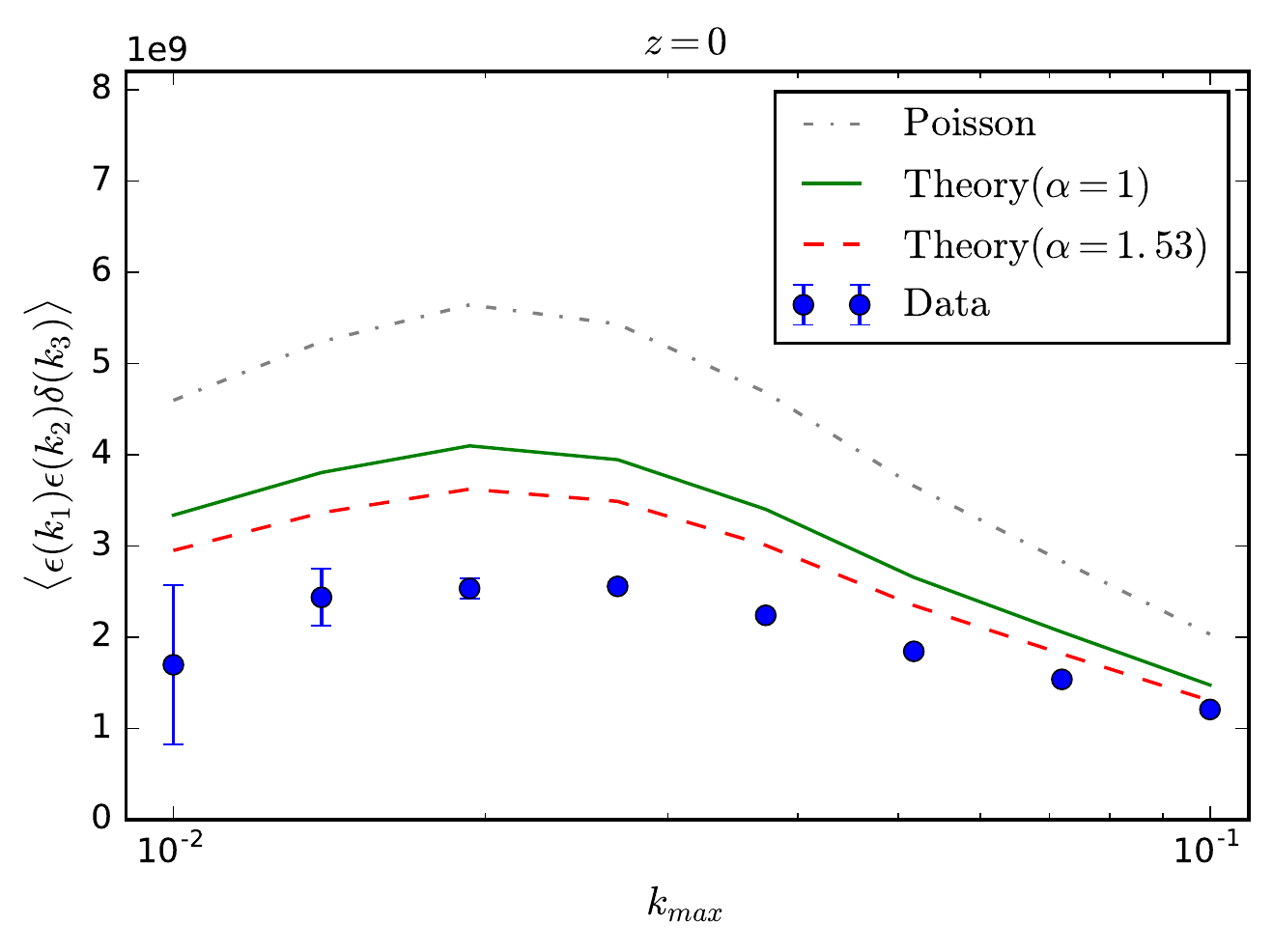}}
\subfloat{\includegraphics[width=0.42\textwidth]{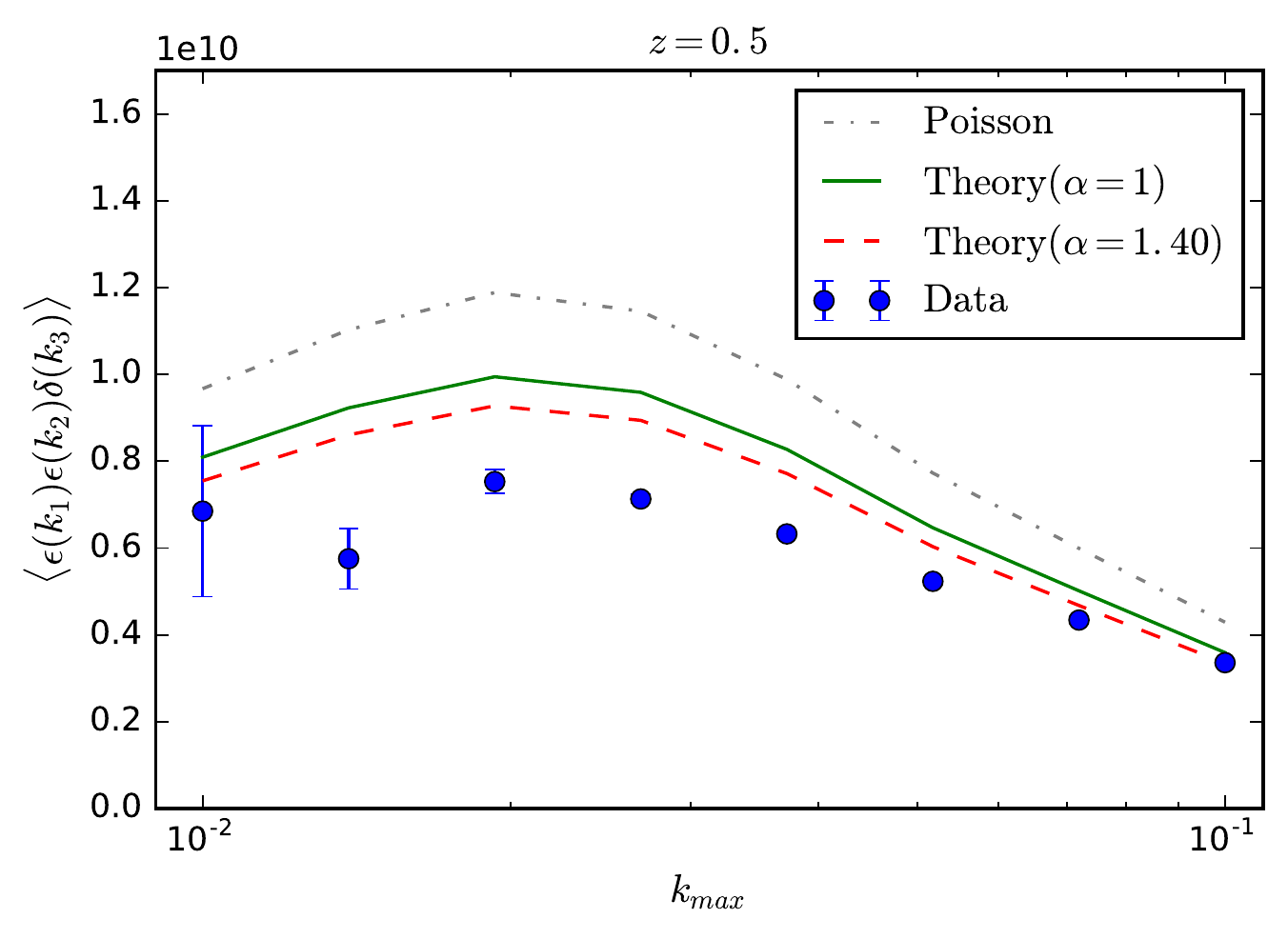}} \\
\subfloat{\includegraphics[width=0.42\textwidth]{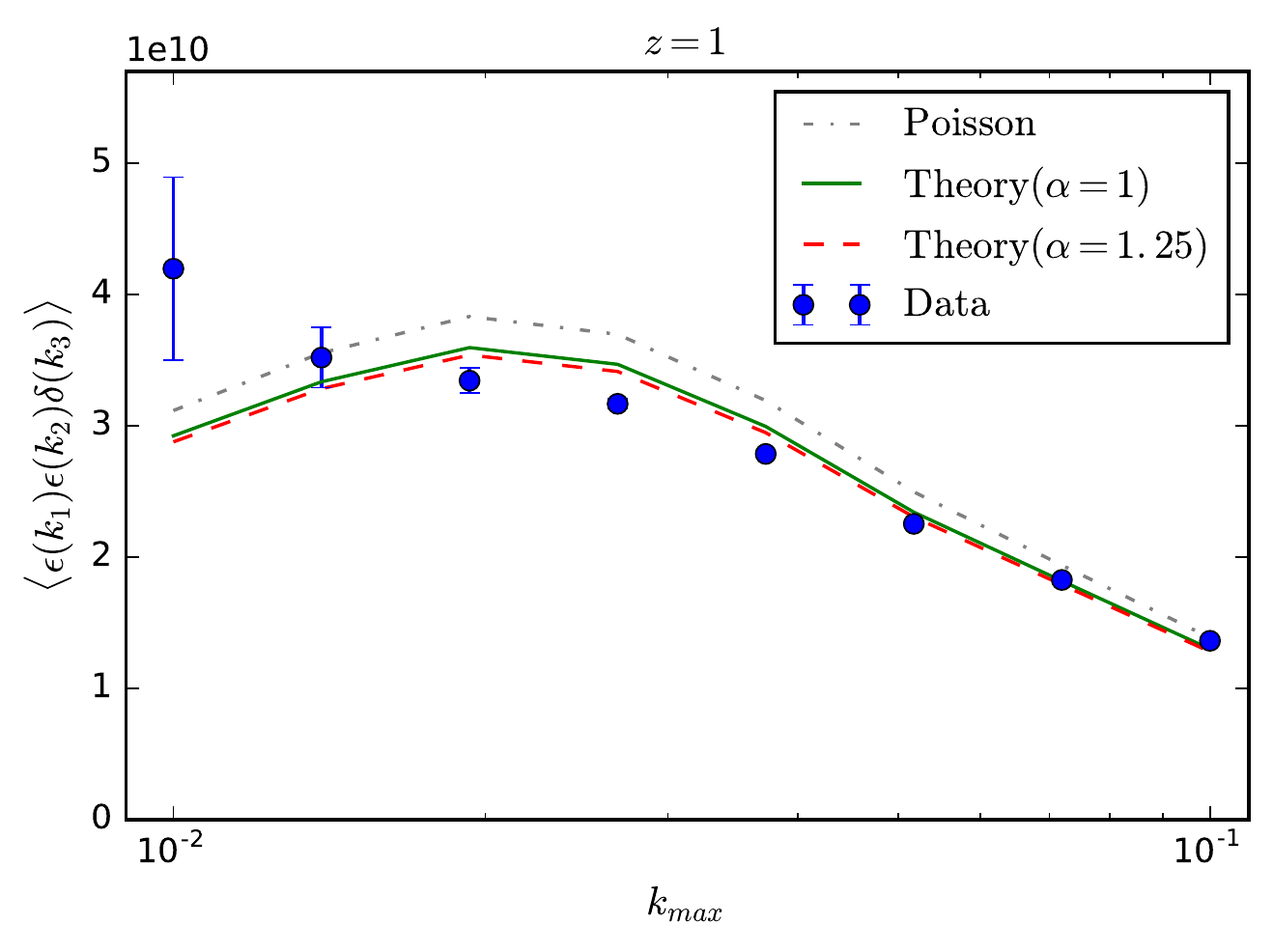}}
\caption{Same as Fig.\ref{fig:Beee}, but for the cummulative cross halo noise - matter bispectrum $B_{\epsilon\epsilon\delta}(<\kmax)$.}
\label{fig:Beed}
\end{figure*}

Finally, Fig.\ref{fig:Beed} displays the measurements of the cumulative cross halo noise matter bispectrum $B_{\epsilon\epsilon\delta}(<\kmax)$ 
which, in the limit of small wavenumber, is given by
\begin{equation}
\label{eq:Be0e0edii}
B_{\epsilon\epsilon\delta}(k_1,k_2,|\vk_1+\vk_2|)=2P_{\epsilon_0\epsilon_{\delta}}^{\{0\}} P_\text{lin}(|\vk_1+\vk_2|) \;.
\end{equation}
This relation is Eq.(\ref{eq:Be0e0ed}) with $i=j$ and $B_{\epsilon_0\epsilon_0\delta}^{\{0\}}\equiv 0$, as predicted by our new halo model 
prescription.
Here, $\vk_1+\vk_2$ is the wavevector carried by $\delta$. To produce a theoretical prediction for this statistics, we compute the cumulative 
power spectrum directly from the data, i.e. we estimate
\begin{equation}
\label{eq:defbmax2}
P_\text{lin}(<\kmax)=\frac{V}{N_t}\sum \big|\delta(\vk_1+\vk_2)\big|^2
\end{equation}
where the sum over the pair $\{\vk_1,\vk_2\}$ is subject to the same aforementioned conditions.
For the amplitude of $P_{\epsilon_0\epsilon_{\delta}}^{\{0\}}$, we consider three estimates: the Poisson expectation 
$P_{\epsilon_0\epsilon_{\delta}}^{\{0\}}=b_1/2\bar{n}$, and the halo model prediction Eq.(\ref{eq:eqswithalpha}) with $\alpha=1$ and $\alpha$
fitted to the noise power spectrum. Overall, the data points consistently trace the shape of the matter power spectrum, in 
agreement with Eq.(\ref{eq:Be0e0edii}). Based on our data only however, we cannot exclude a non-zero value of $B_{\epsilon_0\epsilon_0\delta}^{\{0\}}$
at the level predicted by the ``original'' halo model, i.e. $B_{\epsilon_0\epsilon_0\delta}^{\{0\}}\sim \overline{M}/(\bar n\bar\rho_m) \sim 10^{8-9}$
for the halos considered here. However, as we argued earlier, only $B_{\epsilon_0\epsilon_0\delta}^{\{0\}}\equiv 0$ or, equivalently, 
$\Xi_{ij\delta}\equiv 0$ is consistent with the halo shot noise being uncorrelated with the density.

Here again, the Poisson expectation overestimates the measurements. Our halo model prediction with best-fit $\alpha$ improves the agreement
noticeably, but the match is not perfect. Therefore, the difference between our theoretical predictions and the data certainly cannot
be captured with a single parameter $\alpha$. Nevertheless, the agreement between the various measurements and theoretical predictions is
encouraging, and suggests that it should be possible to understand the shot noise of biased tracers using the halo model prescription advocated 
in this paper.

\section{Discussion}
\label{sec:discussion}

\subsection{Optimal weights}
\label{sec:optimal}

Ref.\cite{Hamaus:2010im} showed that mass weighting reduces the noise of a halo catalogue and returns a weighted halo field whose
correlation with the matter distribution is tighter. This is a consequence of the shot noise 2-point covariance having a zero
eigenvalue, with corresponding eigenvector proportional to the halo mass \cite{Schmidt:2015gwz}. Here, we present yet
another simple derivation of these results. We emphasize that the optimal weight is generally proportional to $M\, n(M)$
unless the halo mass bins are constructed to have identical number density (as is done in \cite{Hamaus:2010im,Hamaus:2011dq}). We 
also point out that mass-weighting not only minimizes the white noise in the weighted power spectrum, but also cancel out the 
weighted bispectrum in the limit where all halos are resolved.

Let us split the halo distribution into mass bins and assign a weight $\lambda_i$ to each of them. For $k\ll 1$, the power spectrum 
of the weighted halo field is given by
\begin{widetext}
\begin{align}
P_{\lambda\lambda}(k) &\equiv 
\sum_{i,j} \lambda_i \lambda_j \left(
b_i b_j P_\text{lin}(k) + \frac{1}{\bar n_i}\delta_{ij}^K 
-b_i \frac{\overline{M_j}}{\bar\rho_m}-b_j\frac{\overline{M_i}}{\bar\rho_m}+b_i b_j\frac{\la n M^2\ra}{\bar\rho_m^2}
\right) \\
&= \sum_{i,j} \lambda_i \lambda_j 
b_i b_j P_\text{lin}(k) 
+\sum_{i,j} \lambda_i \lambda_j \left(
\frac{1}{\bar n_i}\delta_{ij}^K 
-b_i \frac{\overline{M_j}}{\bar\rho_m}-b_j\frac{\overline{M_i}}{\bar\rho_m}+b_i b_j\frac{\la n M^2\ra}{\bar\rho_m^2}
\right) \nonumber \;.
\end{align}
\end{widetext}
Upon taking the continuous limit, $\overline{M_i}$ can be replaced by $M_i$, $\bar n_i$ by the halo mass function $n$ etc.
We shall determine the weights $\lambda_i$ such that the power spectrum of our weighted halo field returns an unbiased estimate 
of the matter power spectrum. 
This leads the following two constraints:
\begin{equation}
\label{eq:cond1}
\int\! dM\, \lambda(M) \, b_1(M) = 1
\end{equation}
and
\begin{align}
\int&\! dM\,\frac{\lambda^2(M)}{n(M)}
+\int\!dM\int\!dM'\, \lambda(M) \lambda(M') \nonumber \\
&\times 
\left(-b_1(M)\frac{M'}{\bar\rho_m}-b_1(M')\frac{M}{\bar\rho_m}+b_1(M) b_1(M')\frac{\la n M^2\ra}{\bar\rho_m^2}\right) 
\nonumber \\
&=0 \;.
\label{eq:cond2}
\end{align}
To determine $\lambda(M)$, we rely on the peak-background split relation Eq.(\ref{eq:PBSb1}) for the linear halo bias and
the halo model assumpation that all the mass is in halos, that is,
\begin{equation}
\int \! dM\,M\,n(M) = \bar \rho_m \;.
\end{equation}
This implies
\begin{align}
\frac{\partial}{\partial\bar\rho_m}\int\! dM\, M\,  n(M) 
&= \int\!dM\,M\frac{\partial n}{\partial\bar\rho_m} \nonumber \\
&= \int\!dM\,\left(M\frac{n}{\bar\rho_m}\right) 
\frac{\bar\rho_m}{ n}\frac{\partial n}{\partial\bar\rho_m} \nonumber \\
&= 1  \;.
\end{align}
Comparing the last equality with Eq.(\ref{eq:cond1}) and taking advantage of Eq.(\ref{eq:PBSb1}), the desired weight is
\begin{equation}
\label{eq:weight}
\lambda(M) = \frac{n(M) M}{\bar\rho_m} \;.
\end{equation}
In contrast to \cite{Hamaus:2010im,Hamaus:2011dq}, there is an additional multiplicative factor of $n(M)$ because we do not 
enforce the mass bins to have the same number density. Nevertheless, there is no contradiction with their findings since 
they either considered finite mass bins with identical number density or directly weighted the halo density 
$n_h(\vx)=\bar n \delta_h(\vx)$, in which case the factor of $\bar n$ was implicitly accounted for.
Substituting this solution into the second condition Eq.(\ref{eq:cond2}) yields
\begin{widetext}
\begin{align}
\int\!dM\,\frac{1}{n}\left(M\frac{ n}{\bar\rho_m}\right)^2
-\frac{2}{\bar\rho_m}&\int\!dM\,M\frac{n}{\bar\rho_m}\left(\frac{\bar\rho_m}{n}
\frac{\partial n}{\partial\bar\rho_m}\right)\int\!dM'\,M'\left(M'\frac{n}{\bar\rho_m}\right)
+\frac{\la n M^2\ra}{\bar\rho_m^2}
\left[\int \!dM\,M\frac{n}{\bar\rho_m}\left(\frac{\bar\rho_m}{n}
\frac{\partial n}{\partial\bar\rho_m}\right)\right]^2 \nonumber \\
&=\frac{\la n M^2\ra}{\bar\rho_m^2}-\frac{2}{\bar\rho_m^2}
\int\!dM\,M\frac{\partial n}{\partial\bar\rho_m}\int\!dM'\, n(M')\,(M')^2
+\frac{\la n M^2\ra}{\bar\rho_m^2} \nonumber \\
&\equiv 0 \;.
\end{align}
\end{widetext}
This demonstrates that the weight Eq.(\ref{eq:weight}) also gives zero shot-noise in the limit where all halos down to $M=0$ 
are resolved. Therefore, weighting halos with $\bar n M/\bar\rho_m$ yields an unbiased estimate of the linear matter power 
spectrum with zero white noise, in agreement with the numerical experiment of \cite{Hamaus:2010im}.

Our derivation can be easily extended to higher-order statistics. Namely, ignoring any primordial non-Gaussianity, the 
bispectrum of the weighted halo field on large scales can be read off from Eq.(\ref{eq:bispectra}):
\begin{widetext}
\begin{align}
B_{\lambda\lambda\lambda}(k_1,k_2,k_3) &\equiv \sum_{i,j,k}\lambda_i\lambda_j\lambda_k
\bigg\{b_1^{(i)} b_1^{(j)} \bigg[ b_2^{(k)} + 2 b_{K^2}^{(k)} \left(\mu_{23}^2-\frac{1}{3}\right)\bigg] P_\text{lin}(k_2)P_\text{lin}(k_3) 
+ \mbox{(2 cyc.)}\bigg\}+\sum_{i,j,k}\lambda_i\lambda_j\lambda_k B_{\epsilon_{0i}\epsilon_{0j}\epsilon_{0k}}^{\{0\}} \;,
\end{align}
\end{widetext}
with the white noise contribution given by Eq.(\ref{eq:Be0e0e0}). We have labelled the halo mass bins with superscripts
to improve the readability of this equation.
Using the peak-background split relation
\begin{equation}
b_N  = \frac{\bar\rho_m^N}{n}\frac{\partial^N n}{\partial\bar\rho_m^N} 
\end{equation}
for the LIMD bias parameters, the weighted average of $b_N$ reduces to
\begin{align}
\int\!dM\,\lambda(M)\,b_N(M) &= \bar\rho_m^{N-1}\frac{\partial^N}{\partial\bar\rho_m^N}\int\!dM\,M\,n(M) 
\nonumber \\
&= \delta_{N1}^K \;,
\end{align}
i.e. they vanish unless $N=1$. The weighted sum of the tidal shear bias $b_{K_2}$ also vanishes in the limit where all halos are 
resolved because the shear is uncorrelated with the average density $\bar\rho_m$. 
Furthermore, one can easily check that, in the same limit, the optimal weight Eq.(\ref{eq:weight}) cancels out the white noise 
contribution to the weighted bispectrum:
\begin{align}
\sum_{i,j,k}\lambda_i&\lambda_j\lambda_k B_{\epsilon_{0i}\epsilon_{0j}\epsilon_{0k}}^{\{0\}} \\ 
&= - \frac{\la n M^3\ra}{\bar\rho_m^3} +3 \frac{\la n M^3\ra}{\bar\rho_m^3} - 3\frac{\la n M^3\ra}{\bar\rho_m^3} 
+\frac{\la n M^3\ra}{\bar\rho_m^3} \nonumber \\ 
&\equiv 0 \nonumber \;.
\end{align}
Therefore, the mass-weighted halo bispectrum vanishes in the limit where the minimum halo mass resolved is $M_\text{min}=0$.
In other words, mass-weighting returns the {\it linear} matter density field at tree-level in perturbation theory. We have not
investigated whether this cancellation occurs at higher order. 

\subsection{From halos to galaxies}
\label{sec:galaxies}

Our approach can be readily extended to compute shot noise corrections to galaxy clustering statistics. Here, we outline how this
can be done assuming, for simplicity, that galaxies follow Poisson statistics, that is, the subhalos hosting galaxies do not exclude 
each other (see, e.g., \cite{Cacciato:2012ut} for a more realistic treatment). We defer a more detailed study to future work.

For illustrative purposes, we shall focus on the $k\to 0$ white noise correction to the galaxy power spectrum, which we also denote
as $P_{\epsilon_0\epsilon_0}^{\{0\}}$. As a rule of thumb, the galaxy shot noise contributions can be obtained upon making the replacement
\begin{equation}
\frac{\Theta(M,M_i)}{\bar n_i} \to \frac{N_g(M)}{\bar n_g} u_g(k|M) 
\end{equation}
in all halo model expressions. Here, $N_g$ is the number of galaxies in halos of mass $M$, whereas $u_g(k|M)$ is the Fourier transform 
of the average density profile of galaxies residing in halos of mass $M$ . The probability distribution $p(N_g)$ of having exactly $N_g$
galaxies in a halo of mass $M$ characterizes as given halo occupation distribution (HOD) model 
(see, for instance,\cite{Zheng:2007zg,Abramo:2015daa} and references therein). 

In practice, galaxy populations are usually split into central and satellite galaxies, which have different colors and morphologies. 
By definition, there is only zero or one central galaxy per halo, i.e. $N_c=0$ or 1. This implies $\overline{N_c^k}=\overline{N_c}$.
Furthermore, one usually assumes that the existence of satellite galaxies is conditioned on the presence of a central galaxy, 
ie. $N_s = N_c \mathcal{N}_s$. Since $P(N_c,\mathcal{N}_s|M)\equiv P(N_c|M)P(\mathcal{N}_s|M)$, this implies 
$\overline{N_c^k \mathcal{N}_s^l}=\overline{N_c^{ }} \overline{\mathcal{N}_s^l}$. Therefore, the average number of galaxies per halo
of mass $M$ is
\begin{equation}
\big\la N_g\big\lvert M\big\ra = \overline N_c(M) (1+\overline{\mathcal{N}_s}(M)) \;, 
\end{equation}
so that the galaxy number density is given by
\begin{equation}
\bar n_g = \int\! dM\, n(M)\, \overline N_c(M) \big[1 + \overline{\mathcal{N}_s}(M)\big] \;.
\end{equation}
Analogously, the linear LIMD galaxy bias $b_g$ and the average host halo mass $\overline{M}_g$ are
\begin{equation}
b_g = \frac{1}{\bar n_g} \int\! dM\, n(M) b_1(M) \overline{N_c}(M) \big[1+\overline{\mathcal{N}_s}(M)\big]
\end{equation}
and
\begin{equation}
\overline{M}_g \equiv \frac{1}{\bar n_g} \int\!dM\,M n(M) \,\overline{N_c}(M)\big[1+\overline{\mathcal{N}_s}(M)\big] \;.
\end{equation}
Note that $\overline{M}_g$ is strongly weighted towards high masses because of its dependence on the number $\mathcal{N}_s$ of satellite 
galaxies. 
Finally, if the number of satellites follows a Poisson distribution, then $\overline{\mathcal{N}_s^2} = \overline{\mathcal{N}_s}$ also 
holds. Note that, although the total galaxy density profile $u_g(k|M)$ is well approximated by the dark matter profile $u(k|M)$, the 
density profiles $u_c(k|M)$ and $u_s(k|M)$ of central and satellite galaxies are generally different because central galaxies tend to 
reside near the host halo center. Nevertheless, this distinction is irrelevant here as we are only interested in the low $k$ behaviour, 
in which all profiles asymptote to unity.

In analogy with halos, the white noise contribution to the galaxy power spectrum thus is given by
\begin{equation}
P_{\epsilon_0\epsilon_0}^{\{0\}} \stackrel{k\to 0}{=} P_{gg}^\text{1H}(k) - 2 b_g P_{g\delta}^\text{1H}(k) + b_g^2 P_{\delta\delta}^\text{1H}(k) \;.
\end{equation}
The contribution from galaxy pairs sitting in the same halo gives the 1-halo term
\begin{align}
\label{eq:1Hgg}
P^\text{1H}_{gg}(k) &\stackrel{k\to 0}{=} \frac{1}{\bar n_g^2}\int\! dM\,n(M) \,\big\la N_g\cdot N_g\big\lvert M \big\ra  \\
&= \frac{1}{\bar n_g}+\frac{1}{\bar n_g^2} \int\!dM\, n(M) \big\la N_g(N_g-1)\big\lvert M\big\ra \nonumber \\
&=\frac{1}{\bar n_g}+
\frac{1}{\bar n_g^2}\int\! dM\, n(M) \,\overline N_c(M) \nonumber \\
&\qquad \times \Big[2\overline{\mathcal{N}_s}(M) +\overline{\mathcal{N}_s}^2(M)\Big] \nonumber \;.
\end{align}
The first term in the second and last equality of Eq.(\ref{eq:1Hgg}) is the usual Poisson noise (induced by self-pairs), whereas the 
second term in the contribution from pairs inside the same halo. There is not factor of $1/2$ because galaxy pairs are formally 
distinguishable. 
Similarly, the 1-halo galaxy - density contribution is
\begin{align}
P^\text{1H}_{g\delta}(k) &\stackrel{k\to 0}{=} \frac{1}{\bar\rho_m \bar n_g}\int\! dM\,M n(M) \,\big\la N_g\big\lvert M \big\ra \nonumber \\
&= \frac{1}{\bar\rho_m \bar n_g}\int\! dM\,M n(M) \,\overline N_c(M) 
\big[1+\overline{\mathcal{N}_s}(M)\big] \nonumber \\
& = \frac{\overline{M}_g}{\bar\rho_m} \;.
\end{align}
Finally, the 1-halo matter power spectrum is unchanged and equal to $\la n M^2\ra/\bar\rho_m^2$ in the limit $k\to 0$. 
Therefore, the galaxy white noise is
\begin{align}
P_{\epsilon_0\epsilon_0}^{\{0\}} &= \frac{1}{\bar n_g}+
\frac{1}{\bar n_g^2}\int\! dM\, n(M) \,\overline N_c \Big(2\overline{\mathcal{N}_s} +\overline{\mathcal{N}_s}^2\Big)
\nonumber \\
& \qquad - 2 b_g\frac{\overline{M}_g}{\bar\rho_m} +b_g^2 \frac{\la n M^2\ra}{\bar\rho_m^2} \;.
\end{align}
The second line in the right-hand side is the non-Poissonian correction. With our approximations, only the first moment $\overline{N_c}$ 
and $\overline{\mathcal{N}_s}$ need to be specified in order to predict the white noise contribution to the galaxy power spectrum.

\section{Conclusions}
\label{sec:conclusions}

Understanding the shot noise of biased tracers beyond the simple Poissonian approximation becomes increasingly relevant as the statistical
uncertainties are expected to decrease significantly in forthcoming galaxy redshift surveys. In this work, we have shown how the halo model
to large scale structure, which has been widely applied and studied, can be used to provide meaningful predictions for the shot noise 
contributions to halo clustering statistics. In essence, the various sources of shot noise in the halo model must be rearranged such that 
they are all absorbed in the halo fluctuation field. The resulting perturbative expansion can be mapped onto the general bias expansion, so
that the various (renormalized) shot noise terms can be clearly expressed as combinations of the halo model shot noises. In combination with 
halo occupation distributions, our approach can furnish useful quantitative estimates for the shot noise contributions to galaxy clustering 
(which remain thus far poorly constrained). 

Furthermore, we have demonstrated how the constant white noise contributions are connected to volume integrals over halo correlation functions
and their response to a long-wavelength density perturbation. These relations define a new class of consistency relations for discrete large
scale structure tracers. Interestingly, our revised halo model precisely reproduces these consistency relations. 
Note that these volume integrals could also be computed using fully nonlinear description of halo clustering, such as peak theory for instance 
(as in, e.g., \cite{Baldauf:2013hka,Baldauf:2015fbu}). However, our halo model prescription has the advantage of being far more flexible as it 
does not require any sophisticated numerical evaluations.

Finally, we have also presented a comparison of our theoretical predictions with measurements of halo shot noise bispectra extracted from 
a large suite of numerical simulations. We have found that, for the massive halos considered here, the various bispectrum shot noise 
statistics are significantly sub-Poissonian, in agreement with \cite{CasasMiranda:2002on,Seljak:2009af} and the more recent power spectrum 
analysis of \cite{Hamaus:2010im}. Our halo model -based predictions fare reasonably well, yet the match is not perfect. Among the possible
sources of discrepancy is the fact that we have assumed a unique, perfectly symmetric halo profile while density profiles certainly vary
in shape on a halo-by-halo basis. Moreover, our assumption $\la\epsilon\delta\ra=0$ etc. implies that the halo profile (which creates the 
exclusion volume responsible for sub-Poissonian noise at large halo mass) is uncorrelated with density fluctuations. This may not hold at 
all order if halo profiles retain memory of the initial conditions or the halo assembly history.

Thus far, our halo model prescription is, arguably, more an educated guess than a rigorous theoretical construction, but we believe it can 
furnish useful insights towards a comprehensive description of the clustering of discrete large scale structure tracers.

\section*{Acknowledgments}

We thank Linda Blot for help with the simulations.
It is a pleasure to acknowledge Nico Hamaus for a careful reading and comments on the manuscript; and 
the participants of the BCCP \& CCA workshop on ``The Nonlinear Universe'' (\v Smartno, Slovenia) for helpful feedback.
This research was supported by the Israel Science Foundation (grant no. 1395/16).

\appendix

\section{A collection of halo model formulae}
\label{app:HMformulae}

We give explicit expressions for the 1-halo, 2-halo and 3-halo contributions to cross halo-matter power spectra and bispectra without
restriction on the wavenumbers. We adopt the notation of \cite{Smith:2006ne,Hamaus:2010im}: each halo fluctuation field brings a factor of 
$\Theta(M,M_i)/\bar n_i$, whereas each matter fluctuation fields carries a factor of $(M/\bar\rho_m) u(k|M)$. While there is no factor of 
$u(k|M)$ for halos (since, by definition, halo centers lie at the center of the average halo profile), matter fluctuation field carry a 
factor of $u(k|M)$. The 1-halo power spectra are given by
\begin{align}
P^\text{1H}_{ij}(k) &= \frac{1}{\bar n_i\bar n_j}\int\!dM\,n(M) \Theta(M,M_i) \Theta(M,M_j)\nonumber \\
P^\text{1H}_{i\delta}(k) &=\frac{1}{\bar n_i\bar\rho_m}\int\!dM\,M n(M) \Theta(M,M_i) u(k|M) \nonumber \\
P^\text{1H}_{\delta\delta}(k) &= \frac{1}{\bar\rho_m^2}\int\!dM\,M^2 n(M) u(k|M)^2 \;,
\label{eq:P1Hk}
\end{align}
where the Fourier transform $u(k|M)$ of the halo profile asymptotes to unity in the large scale limit $k\to 0$.
Analogously, the 1-halo bispectra are given by
\begin{align}
B^\text{1H}_{ijk}(k_1,k_2,k_3) &= \frac{1}{\bar n_i\bar n_j\bar n_k}
\int\!dM\,n(M) \Theta(M,M_i) \Theta(M,M_j) \nonumber \\ 
&\qquad \times \Theta(M,M_k) \nonumber \\
B^\text{1H}_{ij\delta}(k_1,k_2,k_3) &= \frac{1}{\bar n_i\bar n_j\bar\rho_m}
\int\!dM\,M n(M) \Theta(M,M_i) \nonumber \\
&\qquad \times \Theta(M,M_j) u(k_3|M) \nonumber \\
B^\text{1H}_{i\delta\delta}(k_1,k_2,k_3) &= \frac{1}{\bar n_i\bar\rho_m^2}
\int\!dM\,M^2 n(M) \Theta(M,M_i) \nonumber \\
&\qquad \times \prod_{\alpha=2}^3 u(k_\alpha|M) \nonumber \\
B^\text{1H}_{\delta\delta\delta}(k_1,k_2,k_3) &= 
\frac{1}{\bar\rho_m^3}\int\!dM\,M^3 n(M) \prod_{\alpha=1}^3 u(k_\alpha|M) \;.
\label{eq:B1Hk}
\end{align}
Finally, the 2-halo contribution to the bispectra are
\begin{widetext}
\begin{align}
\label{eq:B2Hk}
B^\text{2H}_{ij\delta}(k_1,k_2,k_3) &=
\frac{1}{\bar n_i\bar n_j\bar\rho_m}\int\!dM\,\Theta(M,M_i) n(M) b_1(M)\int\!dM'\,\Theta(M',M_j)M'n(M')  b_1(M') u(k_3|M')
P_\text{lin}(k_1) \\
&\quad + \frac{1}{\bar n_i\bar n_j\bar\rho_m}\int\!dM\,\Theta(M,M_j) n(M) b_1(M) \int\!dM'\,\Theta(M',M_i)M' n(M') b_1(M') u(k_3|M')
P_\text{lin}(k_2)\nonumber \\
& \quad + \frac{1}{\bar n_i\bar n_j\bar\rho_m}\int\!dM\,M n(M) b_1(M)u(k_3|M)\int\!dM'\,\Theta(M',M_i)\Theta(M',M_j) n(M') b_1(M')
P_\text{lin}(k_3) \nonumber \\
B^\text{2H}_{i\delta\delta}(k_1,k_2,k_3) &=
\frac{1}{\bar n_i\bar\rho_m^2}\int\!dM\,\Theta(M,M_i) n(M) b_1(M)\int\!dM'\,(M')^2 n(M')b_1(M') u(k_2|M) u(k_3|M) P_\text{lin}(k_1)
\nonumber \\
&\quad + \bigg\{
\frac{1}{\bar n_i\bar\rho_m^2}\int\!dM\,M n(M) b_1(M)u(k_2|M)\int\!dM'\,\Theta(M',M_i)M'n(M')b_1(M')u(k_3|M')P_\text{lin}(k_2)
+2 \leftrightarrow 3\bigg\}
\nonumber \\
B^\text{2H}_{\delta\delta\delta}(k_1,k_2,k_3) &=
\frac{1}{\bar\rho_m^3}\int\!dM\,M n(M) b_1(M)u(k_3|M)\int\!dM'\,\big(M'\big)^2 n(M')b_1(M')u(k_1|M') u(k_2|M') P_\text{lin}(k_3) + 
\mbox{(2 cyc.)} \nonumber
\end{align}
\end{widetext}
Notice the factor of $b_1(M) b_1(M')$, which follows from expanding the power spectrum of halo centers at linear order.
In the limit $k_i\to 0$, these relations yield Eqs.(\ref{eq:P1H}), (\ref{eq:B1H}) and (\ref{eq:B2H}) of the main text.

\section{The halo shot noise trispectrum}
\label{app:trispec}

To illustrate the applicability of the halo model approach beyond the bispectrum, we compute the Poissonian and non-Poissonian contributions
to the constant white noise in the trispectrum of the halo shot noise $\epsilon_i(\vk)$ (defined in Eq.(\ref{eq:defeps})). We show that this
result furnishes an estimate for the volume integral of the halo 4-point function.

Following our definition of the bispectrum, the trispectrum of 4 fluctuation fields reads
$T_{WXYZ}(k_1,k_2,k_3,k_4) \equiv \langle W(\vk_1) X(\vk_2) Y(\vk_3) Z(\vk_4)\rangle_c '$. Here, the prime signifies that we have removed a 
factor of $(2\pi)^3 \delta^D\!(\vk_1+\vk_2+\vk_3+\vk_4)$ owing to the invariance under translations; whereas the subscript ``c'' signifies
that we should extract the connected piece. To compute the halo noise trispectrum
$T_{\epsilon_i\epsilon_j\epsilon_k\epsilon_l}(k_1,k_2,k_3,k_4)\equiv \langle\epsilon_i(\vk_1)\epsilon_j(\vk_2)\epsilon_k(\vk_3)\epsilon_l(\vk_4)\rangle_c '$, 
we start from the configuration space correlation of four halo noise fields:
\begin{widetext}	
\begin{align}	
\label{eq:Teee}
\big\la\epsilon_i\epsilon_j\epsilon_k&\epsilon_l\big\ra = 
\Big\la\big(\delta_i-b_i\delta\big)\big(\delta_j-b_j\delta\big) \big(\delta_k-b_k\delta\big)\big(\delta_l-b_l\delta\big)\Big\rangle \\
&=\langle\delta_i\delta_j\delta_k\delta_l\rangle
-\Big[b_l\langle\delta_i\delta_j\delta_k\delta\rangle + \mbox{(3 cyc.)}\Big] 
+\Big[b_k b_l\langle\delta_i\delta_j\delta^2\rangle + \mbox{(5 perms.)}\Big]
-\Big[b_j b_k b_l\langle\delta_i\delta^3\rangle + \mbox{(3 cyc.)}\Big] \nonumber \\
&\qquad + b_i b_j b_k b_l\langle\delta^4\rangle \nonumber \;, 
\end{align}
where it is understood that they are evaluated at position $\vx_1$, ... , $\vx_4$.
Substituting the definition of the halo fluctuation field Eq.(\ref{eq:delta_i}), the calculation of the halo-matter 4-point functions is immediate 
(see \cite{Chan:2016ehg} for instance). After some algebra, we obtain:
\begin{align}
\Big\langle \delta_i(\vx_1) \delta_j(\vx_2) \delta_k(\vx_3) \delta_l(\vx_4)\Big\rangle &=
\xi_{ijkl}(\vr_{12},\vr_{13},\vr_{14})+
\bigg[\frac{\delta_{ij}^K}{\bar n_i}\delta^D\!(\vr_{12})\Big(\xi_{kl}(\vr_{34})+\xi_{jkl}(\vr_{23},\vr_{34})\Big) + \mbox{(5 perms.)}\bigg] \nonumber\\
&\qquad +\bigg[\frac{\delta_{ij}^K}{\bar n_i} \frac{\delta_{kl}^K}{\bar n_k}
\delta^D\!(\vr_{12})\delta^D\!(\vr_{34})\Big(1+\xi_{ik}(\vr_{13})\Big) + \mbox{(2 cyc.)}\bigg] \nonumber \\ 
&\qquad +  \bigg[\frac{\delta_{ijk}^K}{\bar n_i^2}\delta^D\!(\vr_{13})\delta^D\!(\vr_{13})\xi_{kl}(\vr_{34})  + \mbox{(3 cyc.)}\bigg] 
+\frac{\delta^K_{ijkl}}{\bar n_i^3} \delta^D\!(\vr_{14})\delta^D\!(\vr_{13})\delta^D\!(\vr_{12}) \nonumber \\
\Big\langle \delta_i(\vx_1) \delta_j(\vx_2) \delta_k(\vx_3) \delta(\vx_4)\Big\rangle &=
\xi_{ijk\delta}(\vr_{12},\vr_{13},\vr_{14})+\bigg[\frac{\delta_{ij}^K}{\bar n_i}\delta^D\!(\vr_{12})\xi_{ik\delta}(\vr_{13},\vr_{14}) + \mbox{(2 cyc.)}\bigg]
+\frac{\delta^K_{ijk}}{\bar n_i^2} \delta^D\!(\vr_{12})\delta^D\!(\vr_{13})\xi_{i\delta}(\vr_{14}) \nonumber \\
\Big\langle \delta_i(\vx_1)\delta_j(\vx_2)\delta(\vx_3)\delta(\vx_4)\Big\rangle &=
\xi_{ij\delta\delta}(\vr_{12},\vr_{13},\vr_{14}) + \frac{\delta_{ij}^K}{\bar n_i} \delta^D\!(\vr_{12}) \xi_{i\delta\delta}(\vr_{13},\vr_{14}) \;.
\label{eq:dddd}
\end{align}
Upon Fourier transforming the result and taking the large scale limit $k_i\to 0$, the Fourier space correlator of four halo fluctuation fields
reads
\begin{align}
\Big\langle \delta_i(\vk_1) \delta_j(\vk_2) \delta_k(\vk_3) \delta_l(\vk_4)\Big\rangle' &\stackrel{k_i\to 0}{=} \Xi_{ijkl}
+\bigg[\frac{\delta_{ij}^K}{\bar n_i}\Big(\delta^D\!(\vk_1+\vk_2) \Xi_{kl}+\Xi_{jkl}\Big) + \mbox{(5 perms.)}\bigg] 
\nonumber \\
&\qquad +\bigg[\frac{\delta_{ij}^K}{\bar n_i}\frac{\delta_{kl}^K}{\bar n_k}
\Big(\delta^D\!(\vk_1+\vk_2)+\Xi_{ik}\Big)+ \mbox{(2 cyc.)}\bigg]
+\bigg[\frac{\delta_{ijk}^K}{\bar n_i^2} \Xi_{kl} + \mbox{(3 cyc.)}\bigg]+ \frac{\delta_{ijkl}^K}{\bar n_i^3} \;.
\label{eq:Tijkl0}
\end{align}
\end{widetext}
Note that we have omitted a factor of $\delta^D\!(\vk_3+\vk_4)$ in the second line, which is trivially implied by invariance under 
translations. Analogously, we find
\begin{align}
\Big\langle \delta_i(\vk_1) \delta_j(\vk_2) \delta_k(\vk_3) \delta(\vk_4)\Big\rangle' &\stackrel{k_i\to 0}{=} 
\Xi_{ijk\delta}+ \frac{\delta_{ij}^K}{\bar n_i} \Xi_{i\delta\delta} \equiv 0
\nonumber\\
\Big\langle \delta_i(\vk_1) \delta_j(\vk_2) \delta(\vk_3) \delta(\vk_4)\Big\rangle' &\stackrel{k_i\to 0}{=} \Xi_{ij\delta\delta}
\equiv 0
\end{align}
for the 4-point cross-covariances since $\Xi$ vanishes as soon as there is at least one density field. Putting these results 
together, and taking advantage of the fact that the connected piece in Eq.(\ref{eq:Tijkl0}) corresponds to the terms without
a Dirac distribution \cite{Matarrese:1997sk}, the white noise contribution 
$T_{\epsilon_i\epsilon_j\epsilon_k\epsilon_l}(k_i\to 0)\equiv T_{\epsilon_{0i}\epsilon_{0j}\epsilon_{0k}\epsilon_{0l}}^{\{0\}}$
to the halo shot noise trispectrum eventually reads
\begin{align}
\label{eq:Te0e0e0e01}
T_{\epsilon_{0i}\epsilon_{0j}\epsilon_{0k}\epsilon_{0l}}^{\{0\}} &= \Xi_{ijkl}+\bigg[\frac{\delta_{ij}^K}{\bar n_i}\Xi_{jkl}+\mbox{(5 perms.)}\bigg] \\
&\quad +\bigg[\frac{\delta_{ij}^K}{\bar n_i}\frac{\delta_{kl}^K}{\bar n_k}\Xi_{ik}+ \mbox{(2 cyc.)}\bigg] \nonumber \\
&\quad +\bigg[\frac{\delta_{ijk}^K}{\bar n_i^2} \Xi_{kl} + \mbox{(3 cyc.)}\bigg]+ \frac{\delta_{ijkl}^K}{\bar n_i^3} \nonumber \;.
\end{align}	
This should be compared to the prediction obtained from our halo model ansatz Eqs.(\ref{eq:newHMnoise}) and (\ref{eq:newPTnoise}).
The contribution arises from the correlators of the halo shot noise $\tilde\epsilon_{0i}$ and $\tilde\epsilon_{0m}$.
Like $B_{\epsilon_0\epsilon_0\epsilon_0}^{\{0\}}$, they are given by the 1-halo trispectrum pieces of the ``original'' halo model 
since the 2-halo etc. terms involve factors of $P_\text{lin}(k)$ etc.
We eventually obtain
\begin{align}
T_{\epsilon_{0i}\epsilon_{0j}\epsilon_{0k}\epsilon_{0l}}^{\{0\}} &=b_i b_j b_k b_l \frac{\la n M^4\ra}{\bar\rho_m^4}-
\bigg[b_i b_j b_k \frac{\overline{M_l^3}}{\bar\rho_m^3}+ \mbox{(3 cyc.)}\bigg] \nonumber \\ 
& \quad +\bigg[b_i b_j\frac{\overline{M_k^2}}{\bar n_k\bar\rho_m^2}\delta_{kl}^K + \mbox{(5 perms.)}\bigg] 
\label{eq:Te0e0e0e02} \\
& \quad - \bigg[b_i\left(\frac{\overline{M_j}}{\bar n_j^2\bar\rho_m}\right)\delta_{jkl}^K + \mbox{(3 cyc.)}\bigg] 
+ \frac{\delta_{ijkl}^K}{\bar n_i^3} 
\nonumber \;.
\end{align}
On comparing Eqs. (\ref{eq:Te0e0e0e01}) and (\ref{eq:Te0e0e0e02}) and using the explicit expressions of $\Xi_{ij}$ and $\Xi_{ijk}$ 
given in Eqs.(\ref{eq:Xij}) and (\ref{eq:Xijk}), the prediction for $\Xi_{ijkl}$ follows immediately.

\bibliographystyle{mn2e}
\bibliography{references}

\begin{thebibliography}{44}
\expandafter\ifx\csname natexlab\endcsname\relax\def\natexlab#1{#1}\fi

\bibitem[{Abramo {et~al}\mbox{.}(2015)Abramo, Balmès, Lacasa, \&
  Lima}]{Abramo:2015daa}
Abramo L.~R., Balmès I., Lacasa F., Lima M., 2015, Mon. Not. Roy. Astron.
  Soc., 454, 2844

\bibitem[{Baldauf {et~al}\mbox{.}(2016)Baldauf, Codis, Desjacques, \&
  Pichon}]{Baldauf:2015fbu}
Baldauf T., Codis S., Desjacques V., Pichon C., 2016, Mon. Not. Roy. Astron.
  Soc., 456, 3985

\bibitem[{Baldauf {et~al}\mbox{.}(2011)Baldauf, Seljak, Senatore, \&
  Zaldarriaga}]{Baldauf:2011bh}
Baldauf T., Seljak U., Senatore L., Zaldarriaga M., 2011, JCAP, 1110, 031

\bibitem[{Baldauf {et~al}\mbox{.}(2013)Baldauf, Seljak, Smith, Hamaus, \&
  Desjacques}]{Baldauf:2013hka}
Baldauf T., Seljak U., Smith R.~E., Hamaus N., Desjacques V., 2013, Phys. Rev.,
  D88, 083507

\bibitem[{Biagetti {et~al}\mbox{.}(2017)Biagetti, Lazeyras, Baldauf,
  Desjacques, \& Schmidt}]{Biagetti:2016ywx}
Biagetti M., Lazeyras T., Baldauf T., Desjacques V., Schmidt F., 2017, Mon.
  Not. Roy. Astron. Soc., 468, 3277

\bibitem[{Biagetti {et~al}\mbox{.}(2013)Biagetti, Perrier, Riotto, \&
  Desjacques}]{Biagetti:2013sr}
Biagetti M., Perrier H., Riotto A., Desjacques V., 2013, Phys. Rev., D87,
  063521

\bibitem[{Blot {et~al}\mbox{.}(2015)Blot, Corasaniti, Alimi, Reverdy, \&
  Rasera}]{Blot:2014pga}
Blot L., Corasaniti P.~S., Alimi J.-M., Reverdy V., Rasera Y., 2015, Mon. Not.
  Roy. Astron. Soc., 446, 1756

\bibitem[{Cacciato {et~al}\mbox{.}(2012)Cacciato, Lahav, Bosch, Hoekstra, \&
  Dekel}]{Cacciato:2012ut}
Cacciato M., Lahav O., Bosch F. C. v.~d., Hoekstra H., Dekel A., 2012, Mon.
  Not. Roy. Astron. Soc., 426, 566

\bibitem[{{Casas-Miranda} {et~al}\mbox{.}(2002){Casas-Miranda}, {Mo}, {Sheth},
  \& {Boerner}}]{CasasMiranda:2002on}
{Casas-Miranda} R., {Mo} H.~J., {Sheth} R.~K., {Boerner} G., 2002, \mnras, 333,
  730

\bibitem[{Chan \& Blot(2016)}]{Chan:2016ehg}
Chan K.~C., Blot L., 2016

\bibitem[{Cooray \& Sheth(2002)}]{Cooray:2002dia}
Cooray A., Sheth R.~K., 2002, Phys. Rept., 372, 1

\bibitem[{Crocce \& Scoccimarro(2008)}]{Crocce:2007dt}
Crocce M., Scoccimarro R., 2008, Phys. Rev., D77, 023533

\bibitem[{de~Putter \& Doré(2017)}]{dePutter:2014lna}
de~Putter R., Doré O., 2017, Phys. Rev., D95, 123513

\bibitem[{Dekel \& Lahav(1999)}]{Dekel:1998eq}
Dekel A., Lahav O., 1999, Astrophys. J., 520, 24

\bibitem[{Desjacques, Jeong \& Schmidt(2016)Desjacques, Jeong, \&
  Schmidt}]{Desjacques:2016bnm}
Desjacques V., Jeong D., Schmidt F., 2016

\bibitem[{Ferraro \& Smith(2015)}]{Ferraro:2014jba}
Ferraro S., Smith K.~M., 2015, Phys. Rev., D91, 043506

\bibitem[{Hamaus, Seljak \& Desjacques(2011)Hamaus, Seljak, \&
  Desjacques}]{Hamaus:2011dq}
Hamaus N., Seljak U., Desjacques V., 2011, Phys. Rev., D84, 083509

\bibitem[{Hamaus, Seljak \& Desjacques(2012)Hamaus, Seljak, \&
  Desjacques}]{Hamaus:2012ap}
Hamaus N., Seljak U., Desjacques V., 2012, Phys. Rev., D86, 103513

\bibitem[{Hamaus {et~al}\mbox{.}(2010)Hamaus, Seljak, Desjacques, Smith, \&
  Baldauf}]{Hamaus:2010im}
Hamaus N., Seljak U., Desjacques V., Smith R.~E., Baldauf T., 2010, Phys. Rev.,
  D82, 043515

\bibitem[{Kehagias \& Riotto(2013)}]{Kehagias:2013yd}
Kehagias A., Riotto A., 2013, Nucl. Phys., B873, 514

\bibitem[{Matarrese, Verde \& Heavens(1997)Matarrese, Verde, \&
  Heavens}]{Matarrese:1997sk}
Matarrese S., Verde L., Heavens A.~F., 1997, Mon. Not. Roy. Astron. Soc., 290,
  651

\bibitem[{Paech {et~al}\mbox{.}(2016)Paech, Hamaus, Hoyle, Costanzi,
  Giannantonio, Hagstotz, Sauerwein, \& Weller}]{Paech:2016hod}
Paech K., Hamaus N., Hoyle B., Costanzi M., Giannantonio T., Hagstotz S.,
  Sauerwein G., Weller J., 2016

\bibitem[{Peacock \& Smith(2000)}]{Peacock:2000qk}
Peacock J.~A., Smith R.~E., 2000, Mon. Not. Roy. Astron. Soc., 318, 1144

\bibitem[{{Peebles}(1980)}]{1980lssu.book.....P}
{Peebles} P.~J.~E., 1980, {The large-scale structure of the universe}

\bibitem[{Pollack, Smith \& Porciani(2012)Pollack, Smith, \&
  Porciani}]{Pollack:2011xp}
Pollack J.~E., Smith R.~E., Porciani C., 2012, Mon. Not. Roy. Astron. Soc.,
  420, 3469

\bibitem[{Rasera {et~al}\mbox{.}(2014)Rasera, Corasaniti, Alimi, Bouillot,
  Reverdy, \& Balmès}]{Rasera:2013xfa}
Rasera Y., Corasaniti P.-S., Alimi J.-M., Bouillot V., Reverdy V., Balmès I.,
  2014, Mon. Not. Roy. Astron. Soc., 440, 1420

\bibitem[{Scherrer \& Bertschinger(1991)}]{Scherrer:1991kk}
Scherrer R.~J., Bertschinger E., 1991

\bibitem[{Schmidt(2016)}]{Schmidt:2015gwz}
Schmidt F., 2016, Phys. Rev., D93, 063512

\bibitem[{Scoccimarro(2000)}]{Scoccimarro:2000sn}
Scoccimarro R., 2000, Astrophys. J., 544, 597

\bibitem[{Scoccimarro {et~al}\mbox{.}(2001)Scoccimarro, Sheth, Hui, \&
  Jain}]{Scoccimarro:2000gm}
Scoccimarro R., Sheth R.~K., Hui L., Jain B., 2001, Astrophys. J., 546, 20

\bibitem[{Scoccimarro, Zaldarriaga \& Hui(1999)Scoccimarro, Zaldarriaga, \&
  Hui}]{Scoccimarro:1999kp}
Scoccimarro R., Zaldarriaga M., Hui L., 1999, Astrophys. J., 527, 1

\bibitem[{Seljak(2000)}]{Seljak:2000gq}
Seljak U., 2000, Mon. Not. Roy. Astron. Soc., 318, 203

\bibitem[{Seljak, Hamaus \& Desjacques(2009)Seljak, Hamaus, \&
  Desjacques}]{Seljak:2009af}
Seljak U., Hamaus N., Desjacques V., 2009, Phys. Rev. Lett., 103, 091303

\bibitem[{Sherwin \& Zaldarriaga(2012)}]{Sherwin:2012nh}
Sherwin B.~D., Zaldarriaga M., 2012, Phys. Rev., D85, 103523

\bibitem[{Sheth \& Tormen(1999)}]{Sheth:1999mn}
Sheth R.~K., Tormen G., 1999, Mon. Not. Roy. Astron. Soc., 308, 119

\bibitem[{Smith(2009)}]{Smith:2008ut}
Smith R.~E., 2009, Mon. Not. Roy. Astron. Soc., 400, 851

\bibitem[{Smith, Desjacques \& Marian(2011)Smith, Desjacques, \&
  Marian}]{Smith:2010fh}
Smith R.~E., Desjacques V., Marian L., 2011, Phys. Rev., D83, 043526

\bibitem[{Smith {et~al}\mbox{.}(2003)Smith, Peacock, Jenkins, White, Frenk,
  Pearce, Thomas, Efstathiou, \& Couchmann}]{Smith:2002dz}
Smith R.~E. {et~al.}, 2003, Mon. Not. Roy. Astron. Soc., 341, 1311

\bibitem[{Smith, Scoccimarro \& Sheth(2007)Smith, Scoccimarro, \&
  Sheth}]{Smith:2006ne}
Smith R.~E., Scoccimarro R., Sheth R.~K., 2007, Phys. Rev., D75, 063512

\bibitem[{Spergel {et~al}\mbox{.}(2007)Spergel {et~al.}}]{Spergel:2006hy}
Spergel D.~N., {et~al.}, 2007, Astrophys. J. Suppl., 170, 377

\bibitem[{Teyssier(2002)}]{Teyssier:2001cp}
Teyssier R., 2002, Astron. Astrophys., 385, 337

\bibitem[{Tinker {et~al}\mbox{.}(2005)Tinker, Weinberg, Zheng, \&
  Zehavi}]{Tinker:2004gf}
Tinker J.~L., Weinberg D.~H., Zheng Z., Zehavi I., 2005, Astrophys. J., 631, 41

\bibitem[{Yoo {et~al}\mbox{.}(2012)Yoo, Hamaus, Seljak, \&
  Zaldarriaga}]{Yoo:2012se}
Yoo J., Hamaus N., Seljak U., Zaldarriaga M., 2012, Phys. Rev., D86, 063514

\bibitem[{Zheng, Coil \& Zehavi(2007)Zheng, Coil, \& Zehavi}]{Zheng:2007zg}
Zheng Z., Coil A.~L., Zehavi I., 2007, Astrophys. J., 667, 760

\end{thebibliography}

\label{lastpage}

\end{document}